\documentclass{aa}  

\usepackage{graphicx}
\usepackage{txfonts}
\usepackage{mathrsfs}  

\def\msun{\hbox{$M_\odot$}}

\def\rg{\hbox{$r_{\rm g}$}}
\def\risco{\hbox{$r_{\rm ISCO}$}}

\def\mdot{\hbox{$\dot{m}_{\rm Edd}$}}
\def\mbh{\hbox{$M_{\rm BH}$}}

\def\rms{\hbox{$r_{\rm ms}$}}
\def\rtr{\hbox{$r_{\rm trans}$}}
\def\rout{\hbox{$r_{\rm out}$}}
\def\spin{\hbox{$a^\ast$}}
\def\ecut{\hbox{$E_{\rm cut}$}}
\def\incl{\hbox{$\theta$}} 

\def\Ltransf{\hbox{$L_{\rm transf} / L_{\rm disc}$}}

\def\Ccal{\hbox{$\mathscr{C}$}}
\def\Gcal{\hbox{$\mathscr{G}$}}

\def\Bcal{\hbox{$\mathscr{B}$}}
\def\Qcal{\hbox{$\mathscr{Q}$}}
\def\kynsed{{\tt KYNSED}}
\def\ntsed{SED$_{\rm NT}$}

\usepackage[breaklinks, colorlinks, citecolor=blue, urlcolor = blue]{hyperref}

\makeatletter
\renewcommand*\aa@pageof{, page \thepage{} of \pageref*{LastPage}}
\makeatother

\begin{document} 

   \title{A physical model for the broadband energy spectrum of X-ray illuminated accretion discs: fitting the spectral energy distribution of NGC~5548}
    \titlerunning{The broadband energy spectrum of X-ray illuminated accretion discs}
    
\author{M. Dov{\v c}iak \inst{\ref{inst1}}
\and
I. E. Papadakis \inst{\ref{inst2},\ref{inst3},\ref{inst1}}
\and
E. S. Kammoun \inst{\ref{inst4}}
\and
W. Zhang \inst{\ref{inst5}}}
\institute{
Astronomical Institute of the Czech Academy of Sciences, Bo{\v c}n{\'i} II 1401, CZ-14100 Prague, Czech Republic \email{michal.dovciak@asu.cas.cz} \label{inst1}
\and
Department of Physics and Institute of Theoretical and Computational Physics, University of Crete, 71003 Heraklion, Greece \label{inst2} 
\and 
Institute of Astrophysics, FORTH, GR-71110 Heraklion, Greece\label{inst3} 
\and 
IRAP, Universit\'e de Toulouse, CNRS, UPS, CNES 9, Avenue du Colonel Roche, BP 44346, F-31028, Toulouse Cedex 4, France\label{inst4} 
\and
National Astronomical Observatories, Chinese Academy of Sciences, 20A Datun Road, Beijing 100101, China\label{inst5}}

\date{Received ; accepted }

 
  \abstract
   {}
   {We develop a new physical model for the broadband spectral energy distribution (SED) of X-ray illuminated accretion discs, that takes into account the mutual interaction of the accretion disc and the X--ray corona, including all the relativistic effects induced by the strong gravity of the central black hole on light propagation and on the transformation of the photon's energy, from the disc to/from the corona rest-frames, and to the observer.}
   {We assume a Keplerian, optically thick and geometrically thin accretion disc, 
   and an X--ray source in the lamp-post geometry. The X--ray corona emits an isotropic, power-law like X-ray spectrum, with a high-energy cut-off. We also assume that all the energy that would be released by thermal radiation in the standard disc model in its innermost part, is transported to the corona, effectively cooling the disc in this region. In addition, we include the disc heating due to thermalisation of the absorbed part of the disc illumination by the X--ray source. X--ray reflection due to the disc illumination is also included. The X--ray luminosity is given by the energy extracted from the accretion disc (or an external source) and the energy brought by the scattered photons themselves, thus energy balance is preserved. We compute the low-energy X--ray cut-off through an iterative process, taking full account of the interplay between the X--ray illumination of the disc and the resulting accretion disc spectrum which enters the corona. We also compute the corona radius, taking into account the conservation of the photon's number during Comptonization.}
   {We discuss in detail the model SEDs and their dependence on the parameters of the system. We show that the disc-corona interaction has profound effects on the resultant SED --- it constrains the X-ray luminosity and changes the shape and normalisation of the UV/optical blue bump. We also compare the model SEDs with those predicted from similar models currently available. We use the new code to fit the broad-band SED of NGC~5548, which is a typical Seyfert~1 galaxy. When combined with the results from previous model fits to the optical/UV time-lags of the same source, we infer a high black-hole spin, an intermediate system inclination, and an accretion rate below 10\% of Eddington. The X-ray luminosity in this source could be supported by 45-70\% of the accretion energy dissipated in the disc. The new model, named \kynsed, is publicly available to be used for fitting AGN SEDs inside the {\tt XSPEC} spectral analysis tool.}
   {X--ray illumination of the accretion disc in AGN can explain {\it both} the observed UV/optical time-lags as well as the broad band SED of at least one AGN, namely NGC~5548. A simultaneous study of the optical/UV/X--ray spectral/timing properties of those AGN with multi-wavelength, long monitoring observations in the last few years, will allow us to investigate the  X--ray/disc geometry in these systems, and constrain their physical parameters.}


   \maketitle
%

\section{Introduction}
\label{sec:intro}

It is widely accepted that active galactic nuclei (AGN) are powered by accretion of matter onto a supermassive black hole (BH). \citet[][hereafter SS73]{Shakura73} and \citet[][hereafter NT73]{Novikov73} are the standard accretion disc models, that have been extensively used to study the accretion process in these objects. These models cannot account for the copious emission of X--rays from AGN. Somehow, through a yet unknown mechanism, accretion power is released to a region where electrons are heated, and then up-scatter the ultraviolet (UV)/optical photons emitted by the disc to X-rays. This region is also known as the X--ray corona. 

The corona size and location are not known at the moment. X--ray monitoring of several lensed quasars has shown that the half-light radius of the X--ray source is less than a few tens of gravitational radii \citep[e.g., see Fig. 1 in][]{Chartas16}. The corona is probably located close to the BH, where most of the accretion power is released. Further information regarding its location with respect to the accretion disc is provided by X--ray spectroscopy. 

More than thirty years ago, {\it GINGA} observations showed that a strong fluorescent iron  K$\alpha$ emission line and a  considerable flattening of the X--ray continuum at energies higher than $\sim 10$ keV were common features in the X--ray spectra of bright Seyferts \citep[e.g.,][]{Nandra91}. Both features were interpreted as evidence for reprocessing of the X-ray continuum by a slab of cold and dense gas \citep[e.g.,][]{Pounds90,George1991}. If this were the accretion disc, then the X--rays should be located above the disc, to be able to illuminate it. 

This scenario makes one implicit prediction. As most of the X--rays will be absorbed, they will generate heat in the disc. As a result, the disc will emit thermal radiation at frequencies which will depend on its effective temperature. We therefore expect the thermally reprocessed radiation to emerge in the UV/optical part of the spectrum. Furthermore, since the X--rays are highly variable in AGN, we would expect the X--ray and UV/optical variations to be correlated, with a delay that should depend on the geometry and the physical properties of the system (accretion rate, X--ray luminosity, etc.). 

Many multiwavelength monitoring campaigns of nearby Seyferts have been performed the last decades in order  to study the correlation between the X-ray/UV/optical variations. The observational effort has significantly intensified the last 10 years. A few AGN have been observed continuously and intensively for many months, over a wide range of spectral bands, from optical to X-rays \citep[e.g.,][]{McHardy14, Fausnaugh16, Cackett18, Edelson19, Cackett20, Pahari20, Santisteban20, Vincentelli21, Kara21}. In all cases, the UV and optical variations are well correlated, but with a delay, which increases from the UV to the optical bands, as expected in the case of X–ray illumination of the disc. The dependence of the time-lags on the wavelength is in agreement with the predictions of the standard accretion disc models. However, the lags amplitude appeared to be larger than expected, by a factor of $\sim 2-3$, given the BH mass and accretion rate of the objects. 

Recently, we studied in detail the response of a standard accretion disc in the UV/optical bands when it is illuminated by X--rays in the lamp-post geometry \citep[][hereafter K21a]{Kammoun21a}. We considered all relativistic effects to determine the incident X-ray flux on the disc, and in propagating light from the source to the disc and to the observer. We computed the disc reflection flux taking into consideration the disc ionization, we calculated the disc response for many model parameters, and we provided a model relation between the X-ray and the UV/optical time-lags, which depends on the  black hole mass, accretion rate, the X-ray corona height and the (observed) 2--10 keV band luminosity (for BH spin 0 and 1). Using this model, we showed that the observed UV/optical time-lags in AGN are fully consistent with the standard accretion disc model, as long as the disc is illuminated by the corona which is located $\sim 30-40$ \rg\, above the BH \citep[][hereafter K21b]{Kammoun21b}.

In this work, we present a new model which can be used to fit the broadband spectral energy distribution (SED) of AGN, from optical/UV to X-rays. The model is based on the K21a model, with some important modifications. K21a considered the X--ray luminosity as a free parameter. In the new model, we also consider the possibility that, through some physical mechanism, all the accretion power below a transition radius, \rtr, is transferred to the corona. In this way, a direct link between the disc and the X--ray corona is established. The X--ray luminosity is not a free parameter any more, but rather depends on the accretion rate and \rtr. In this respect, the model is similar to {\tt AGNSED}, which is a model that can fit the broadband AGN SEDs. {\tt AGNSED} was developed by \cite{Kubota18}, and  it is an update of an earlier model by \cite{Done2012}, called {\tt optxagnf}. Apart from this similarity, there are also many differences between our model and {\tt AGNSED}. We compare the two models in detail in \S~\ref{sec:sed}.

The new model includes a better treatment for the determination of the low-energy cut-off in the X--ray spectrum, which is an important parameter to determine the X--ray flux incident on the disc (and hence the amount of the X-rays that are absorbed and the importance of X--ray illumination), as well as the X--ray spectrum normalization. 
It also considers conservation of the number of the disc photons, as they scatter off the hot electrons in the corona. In this way, the model determines the size of the X--ray corona. This should be compared with the height of the source, as the corona radius should be smaller than the corona height minus the horizon radius, at least. Any discrepancies should indicate that the adopted lamp-post geometry is not consistent with the data.    

In the second part of the paper, we present the results from the model fitting of the average, broadband energy spectrum of NGC~5548. This was one of the first AGN that was simultaneously observed in the X--rays and in the UV with the aim to test whether  the variations in these two bands are correlated or not \citep{Clavel92}. It was also monitored by {\it Swift}, {\it HST} and ground based telescopes over long periods the last few years \citep{McHardy14,Edelson15}. 
In \citet[K19, hereafter]{Kammoun19lags} and K21b, we fitted the observed time-lags spectra of the source reported by \cite{Fausnaugh16}, and we found a very good agreement with a standard disc, which accretes at a few percent of the Eddington limit, and is illuminated by an X--ray source located at a height of $20-60$ or $30-80$~\rg, depending on whether the BH spin is 0 or 1. In addition, we also studied the X--ray, UV and optical power spectra of the {\it STORM} light curves \citep{Panagiotou20}. We used the disc response functions of K19 and we found that the amplitude of the UV and optical PSDs are also compatible with a standard disc being illuminated by X-rays, with low accretion rates. 

The paper is organised as follows: in \S\ref{sec:model} we explain our assumptions and describe all the details of the model set-up, in \S\ref{sec:sed} we show the broadband SED predicted by our model for different values of model parameters, in \S\ref{sec:fitting} we apply the model to the broadband energy spectrum of NGC~5548 and in \S\ref{sec:discussion} we finish by summing up our results and conclusions.

\section{The model set up}
\label{sec:model}

We have already studied the disc thermal reverberation in K19, K21a and K21b. K21a developed a code that can compute the disc response function, for a given BH mass, accretion rate, X--ray luminosity, etc.. In this work, we present a new code, {\tt KYNSED}, which can be used to fit the broadband, spectral energy distribution (SED) of AGN, from optical to hard X-rays. The model is similar to the model outlined in K21a, with some important modifications which we describe in the following sections.

We consider a Keplerian, geometrically-thin and optically-thick accretion disc with an accretion rate of \mdot\footnote{\mdot is the accretion rate normalized to the Eddington accretion rate, i.e.\, \mdot=$\dot{M}/\dot{M}_{\rm Edd}$, where $\dot{M}$ is the accretion rate in physical units, and $\dot{M}_{\rm Edd}$ is the accretion rate for the total disc luminosity to be equal to Eddington luminosity. It depends on the accretion efficiency, i.e., on BH spin.}, around a BH with a mass of $M_{\rm BH}$ and  spin $a^\ast$.\footnote{\spin$= Jc/G$\mbh$^2$, where $J$ is the angular momentum of the BH, and is smaller or equal to 1, e.g., \cite{Misner73}.} 
We assume that the intrinsic disc temperature profile follows the NT73 prescription. We take into account the spectral hardening due to photon interactions with matter in the upper disc layers via
a colour correction factor, $f_{\rm col}$, as described by \citet{Done2012}. It depends on the effective temperature of the disc, hence on all parameters that determine the disc temperature (e.g., \mbh, \mdot, X--ray incident flux etc). Following \citet{Done2012}, we assume that $f_{\rm col}=1$ when $T_{\rm d}(r)<3\times 10^4\,$K, then $f_{\rm col}=(T_{\rm d}(r)/3 \times 10^4\,{\rm K})^{0.82}$ over the critical temperature range of $3\times 10^4\,{\rm K} < T_{\rm d}(r) < 10^5\,{\rm K}$, and then $f_{\rm col}=(72\,{\rm keV}/k_{\rm B}T_{\rm d}(r))^{1/9}$ at higher disc temperatures ($T_d(r)$ is the disc temperature in kelvins, which is set by the energy released locally by the accretion process and by the X-ray absorbed flux, and $k_{\rm B}=8.62\times10^{-8}\,{\rm keV/K}$ is Boltzmann constant). The colour correction factor changes the disc temperature and the normalisation of the emitted black body radiation at each radius, $r$, in such a way that the total power radiated does not change. 

As for the X--ray source, we assume the lamp-post geometry: the X-ray corona is point-like, located at a height $h$ above the BH, on its rotational axis. We also assume that the X--rays are emitted isotropically in the rest-frame of the corona. They illuminate the disc, which will increase its temperature due to the thermalisation of the absorbed incident flux.


As K21a showed (see their eq. (4), for example), the observed disc flux at a particular wavelength, $F_{\rm obs}(\lambda, t)$, will be equal to the NT disc flux, $F_{\rm NT}(\lambda)$, plus an extra term due to the disc heating by the X-ray absorption, i.e.

\begin{equation}\label{eq:Fobs}
    F_{\rm obs}(\lambda, t) = F_{\rm NT}(\lambda) + \int_{0}^{+\infty} L_{\rm X}(t-t')\Psi_{L_{\rm X}}(\lambda, t'){\rm d}t'.
\end{equation}

\noindent The equation above assumes that the disc variability is due only to the (variable) X-ray illumination of the disc. The convolution in the right hand side gives the thermally reprocessed disc emission. The term $L_{\rm X}$ in K21a is equal to the observed, $2-10$~keV luminosity of the corona (in Eddington units). In general, it can be any quantity representative of the corona luminosity. For example, it can be the total X--ray luminosity itself, as long as the disc response is normalized accordingly. $\Psi_{L_{\rm X}} (\lambda, t)$ is the disc response function at $\lambda$,  which depends on the X--ray luminosity in a non linear way (see K21a). 


It is difficult to model the observed AGN SEDs using eq.\,(\ref{eq:Fobs}). If we use simultaneous observations to construct the optical/UV SED, we would need continuous X-ray observations over a large period of time in order to compute the convolution integral in the right part of eq.\,(\ref{eq:Fobs}). The duration of the X--ray observations  depends on the X--ray variability amplitude of each source and on the width of the disc response function, which is wavelength dependent. For example, modelling the contemporaneous SED of an AGN with measurements from $\sim 1000$ up to $\sim 10000$~\AA\ will require continuous X-ray observations over $\sim ~1$ up to $\sim 20-30$ days prior to the optical/UV observations, to compute the convolution integral in eq.\,(\ref{eq:Fobs}) for the shortest and the longest wavelength in the SED \footnote{The numbers are indicative, and correspond to the time at which the respective disc response function decreases by a factor of a few, as compared to its maximum value. The numbers correspond to the 1158\AA\ and the $z-$band disc response plotted in the top panel in Figure\,8 in K21a. They depend on the wavelength, and on the model parameters considered.}. 

Equation (\ref{eq:Fobs}) can be used to study the time average SED since,

\begin{equation}\label{eq:Fobs2}
    \langle F_{\rm obs}(\lambda,t)\rangle = F_{\rm NT}(\lambda) + \langle \int_{0}^{\infty} L_{\rm X}(t-t') \Psi_{L_{\rm X}}(\lambda, t'){\rm d}t'\rangle,
\end{equation}

\noindent where brackets denote the mean. 
K21a showed (see their Appendix A) that, even if X--rays vary by a factor of $\sim 10$ or so, the integral in the right hand side of eq.\,(\ref{eq:Fobs2}) will be almost equal to the product of the mean X--ray luminosity, $\langle L_{\rm X}\rangle$ and the integral of $\Psi_{\langle L_{\rm X}\rangle}(t)$, i.e., the  response function that corresponds to $\langle L_{\rm X}\rangle$. In this case, eq.\,(\ref{eq:Fobs2}) can be re-written as

\begin{equation}\label{eq:Fobs3}
    \langle F_{\rm obs, t}(\lambda,t) \rangle = F_{\rm NT}(\lambda) + \langle L_{_{\rm X}} \rangle \int_{0}^{\infty}  \Psi_{\langle L_{\rm X}\rangle}(\lambda, t){\rm d}t.
\end{equation}

\noindent \kynsed\ uses the above equation to compute the mean X--ray/UV/optical SED in AGN. We explain in detail the model in the following Sections. 

\subsection{The transfer of the energy from the accretion disc to the corona} \label{sec:energy_transfer}

The power that heats the corona is a free parameter in K21a. In this work, we also consider the possibility that the power from the accretion flow 
that is used to heat the gas below a transition radius\footnote{\rtr, as well as all other radii, are measured with respect to the gravitational radius, \rg$=G$\mbh$/c^2$.}, \rtr, is transferred to the corona by an unknown physical mechanism \citep[i.e., similarly as was done in {\tt AGNSED}, see][]{Kubota18}. In this way the luminosity of the X--ray source is not a free parameter anymore: \rtr\ determines the total luminosity of the corona. 

We also assume that the accretion disc is still Keplerian below \rtr, with unchanged radial transfer of energy and angular momentum. If the power that is transferred to the corona, $L_{\rm transf}$, is equal to the disc luminosity (at infinity) emitted below the transition radius, then  

\begin{equation}
L_{\rm transf}=2\pi \int_{\rms}^{\rtr}\,\sigma T^4_{\rm NT}(r) [-U_t(r)]\,r\,{\rm d}r,
\label{eq:ltransf}
\end{equation}

\noindent where \rms\, is the radius of the marginally stable orbit (also referred to as the radius of the innermost circular stable orbit, ISCO, \risco), $-U_t(r)$ is the time component of the accretion disc 4-velocity that transforms the locally released accretion energy to the one with respect to the observer at infinity, and $T_{\rm NT}(r)$ is the radially dependent NT73 effective disc temperature. The integral of the equation above over the whole accretion disc gives the total disc luminosity, 

\begin{equation}
L_{\rm disc}=2\pi\int_{\rms}^{\infty}\,\sigma T^4_{\rm NT}(r)[-U_t(r)]\,r\,{\rm d}r=\eta\dot{M}
c^2,
\end{equation}

\noindent where $\dot{M}$ is the disc accretion rate, and $\eta=1+U_t(\rms)$ is the accretion efficiency. 
The fraction of the power extracted from the accretion flow below \rtr\, over the 
total disc luminosity is given by \citep[here, we used the equation for the energy balance at a given radius, see e.g., eq.~(3.171) in][]{Kato98},

\begin{equation}
\frac{L_{\rm transf}}{L_{\rm disc}} = \frac{\Ccal^{-1/2}\Gcal - r^{-1}_{\rm trans}\,\Bcal^{-1}\Qcal + U_t(\rms)}{1+U_t(\rms)},
\label{eq:ftransf}
\end{equation}

\noindent where \Bcal, \Ccal, \Gcal\/ and \Qcal\/ are relativistic correction factors defined, for example in \cite{Page1974}, as functions of \rtr, and

\begin{equation}
U_t(\rms) = -\frac{ 4\,(\sqrt{\rms} - \spin) + \spin}{\sqrt{3}\rms},
\end{equation}

\noindent is the 4-velocity of the accreting matter at ISCO (we have used 
eq. (5.4.8) in NT73 to evaluate $U_t(\rms)$ in this simple form).

\begin{figure}
\centering
\includegraphics[width=0.9\linewidth]{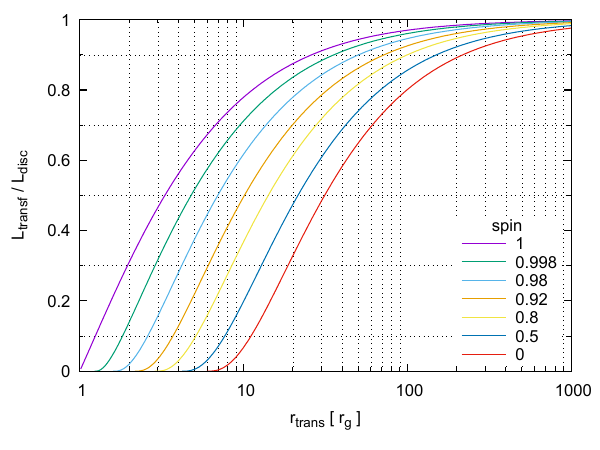}
\caption{The dependence of 
\Ltransf\, on the transition radius, \rtr. 
}
\label{fig:Ltransf_spin}
\end{figure}

The dependence of the fraction, \Ltransf, on the transition radius, \rtr, is shown in Figure~\ref{fig:Ltransf_spin} for several values of \spin. The figure shows that it rises very quickly to one, as expected, since most of the accretion power is released in the inner disc. Half of the total accretion 
luminosity of the standard disc is released within $\sim 3$~\rg\, and $\sim 30$~\rg, when $\spin=1$ and $\spin=0$, respectively. If the transition radius is at this distance from the center then, in our model, almost half of the accretion 
luminosity of the standard disc will be transferred to the corona instead of being radiated away. 

The power extracted from the accretion flow evaluated in the reference frame of the corona will then be
\begin{equation}
L_{\rm transf, c}= \frac{1}{2}(U^t_{\!{\rm c}})^2 \left(\frac{L_{\rm transf}}{L_{\rm disc}}\right) \eta\dot{M}_{\rm BH} c^2,
\label{eq:Ltransf_cor}
\end{equation}

\noindent where \Ltransf\, is given by eq.~(\ref{eq:ftransf}) and $U_{\!{\rm c}}^t(h,\spin)=[1-2h/(h^2+\spin^2)]^{-1/2}$ accounts for the transformation of the extracted power at infinity into the local frame of corona. The subscript ``c'' in eq.~(\ref{eq:Ltransf_cor}), and throughout the paper, indicates quantities that are evaluated at the corona reference frame. The factor 1/2 accounts for the fact that only half of the extracted energy from the accretion flow will be used to heat the corona above the accretion disc, towards the direction of the observer, the other half being used for the corona on the other disc side.

\subsection{The X--ray energy spectrum and the corona size}
\label{sec:2p2}

We assume that the X-ray corona emits isotropically (in its rest frame) a power-law photon flux of the form,

\begin{equation}
\label{eq:Xrayspectrum}
f_{\rm X, c}(E)\equiv \frac{{\rm d}N}{{\rm d}t{\rm d}\Omega{\rm d}E} = A_{\rm c}\,E^{-\Gamma}\exp(-E/E_{\rm cut,c}),
\end{equation}

\noindent where $A_{\rm c}$ is the power-law normalisation, $\Gamma$ is the spectral photon index (assumed to be constant), and $E_{\rm cut,c}$ is the exponential cut-off at high energies. The X--ray spectrum extends down to a low energy rollover, $E_{\rm 0,c}$.
$\Gamma$ and $E_{\rm cut,c}$ are free parameters of the model, while $A_{\rm c}$ and $E_{\rm 0,c}$ are determined by the other model parameters, as we explain below. 

The value of $E_{\rm 0,c}$ is determined by the average energy of the seed photons, as seen by the corona,

\begin{equation}
\label{eq:E0}
E_{\rm 0,c} = \frac{L_{\rm BB, c}}{N_{\rm BB, c}},
\end{equation}

\noindent where $L_{\rm BB, c}$ and $N_{\rm BB, c}$ are the luminosity and photon flux of the thermal radiation from the accretion disc illuminating the corona, both evaluated per unit area at the corona rest frame (see the following section for their estimation). The normalization is determined as follows, 

\begin{equation}
\label{eq:Xraynorm}
A_{\rm c}=\frac{L_{\rm X, c}}{4\pi\,\Gamma_{\rm f}(2-\Gamma,E_{\rm 0,c}/E_{\rm cut,c})},
\end{equation}

\noindent  where $\Gamma_{\rm f}$ is the incomplete Gamma function and,

\begin{equation}
\label{eq:Lx0}
L_{\rm X, c} =  4\pi \int_{E_0}^{\infty} Ef_{\rm X,c}(E)\,{\rm d}E = 
L+(1-{\rm e}^{-\tau})\,{\rm \Delta}S_{\!{\rm c}}\,L_{\rm BB, c},
\end{equation}

\noindent where $L$ is the power given to the corona. It is either equal to $L_{\rm transf,c}$ or, if the corona is powered by some external mechanism that is not related to the accretion process, then it is a free parameter, say $L_{\rm ext,c}$. The equation (\ref{eq:Lx0}) states the conservation of energy: the total luminosity of the X--ray photons, $L_{\rm X, c}$, is equal to the sum of the power that is used to heat the electrons in the corona plus the energy of the photons that entered the corona and were scattered. The optical depth of the corona, $\tau$, can be evaluated from the spectral slope $\Gamma$ and the corona temperature. $\Gamma$ is a free model parameter, and we assume an electron temperature of 100~keV for the estimation of $\tau$, like in \citet[][ hereafter DD16]{Dovciak2016}. In the above equation, $\Delta S_{\!{\rm c}}$ is the local cross-section of the corona and depends on the corona radius, $R_{\rm c}$, given in Boyer-Lindquist coordinates, as follows (DD16): 

\begin{equation}
\label{eq:deltas}
\Delta S_{\!{\rm c}} = U_{\!{\rm c}}^t\,\pi R_{\rm c}^2.
\end{equation}

\noindent We cannot determine $L_{\rm X,c}$ from eq.~(\ref{eq:Lx0}), as $\Delta S_{\rm c}$ is not known. We therefore considered the fact that the Comptonization process conserves the photon number. This implies that

\begin{equation}
\label{eq:nx0}
N_{\rm c}  =  4\pi\int_{E_0}^{\infty} f_{\rm X, c}(E)\,{\rm d}E = (1-{\rm e}^{-\tau})\,\Delta S_{\!{\rm c}}\,N_{\rm BB, c},
\end{equation}

\noindent where $N_{\rm c}$ is the number of the X--ray photons emerging from the corona. The equations (\ref{eq:Lx0}) and (\ref{eq:nx0}) include only two unknown parameters, namely $A_{\rm c}$ (hence $L_{\rm X,c}$) and $\Delta S_{\rm c}$ (hence, the corona radius). Solving them eventually gives us, 

\begin{eqnarray}
\label{eq:Lx}
L_{\rm X, c} & = & \frac{L}
{1-E_{\rm 0,c}/E_{\rm sc,c}}\ ,\\
\label{eq:Rc}
R_{\rm c} & = & \left[\frac{L}
{\pi\,(1-{\rm e}^{-\tau})\,U_{\!{\rm c}}^t\,(E_{\rm sc,c}-E_{\rm 0,c})\,N_{\rm BB, c}}\right]^{1/2}\ ,
\end{eqnarray}

\noindent where, 

\begin{equation}
\label{eq:Ec}
E_{\rm sc,c} =  \frac{L_{\rm X, c}}{N_{\rm c}} = \frac{\Gamma_{\rm f}(2-\Gamma, E_{\rm 0,c}/E_{\rm cut,c})}{\Gamma_{\rm f}(1-\Gamma, E_{\rm 0,c}/E_{\rm cut,c})}\,E_{\rm cut,c},
\end{equation}

\noindent is the average energy of the outgoing, scattered X-ray photons. Note that the equations (\ref{eq:Lx}) and (\ref{eq:Rc}) are valid for both cases when $L=L_{\rm transf,c}$ or $L=L_{\rm ext,c}$.

\subsection{The thermal flux of the accretion disc received by the corona}
\label{sec:2p3}

As we showed above, the intrinsic X-ray luminosity, as well as the corona radius, depend on the disc thermal luminosity, $L_{\rm BB, c}$, and photon flux, $N_{\rm BB, c}$, evaluated at the corona rest frame. The disc spectrum at the corona is that of a multi-colour black body. The luminosity and photon number density per unit area that the corona receives from the disc, when accounting for all relativistic effects, is (see also eqs.~(1) and (2) in DD16),

\begin{eqnarray}
\label{eq:Lbb}
L_{\rm BB, c} & = & 2\sigma\int_{\rms}^{\rout}\frac{T^4_{\!{\rm d}}(r)}{g^4(r)f^4_{\rm col}(r)}\,\frac{{\rm d}\Omega_{\rm c}}{{\rm d}S}(r)\,r\,{\rm d}r\ ,\\
\label{eq:Nbb}
N_{\rm BB, c} & = & 2\sigma_{\rm p}\int_{\rms}^{\rout}\frac{T^3_{\!{\rm d}}(r)}{g^3(r)f^4_{\rm col}(r)}\,\frac{{\rm d}\Omega_{\rm c}}{{\rm d}S}(r)\,r\,{\rm d}r,\
\end{eqnarray}

\noindent  where $f_{\rm col}(r)$ is the colour correction factor, and $T_{\rm d}(r)$ is the disc temperature (see \S\ref{sec:2p4}). The integration over the whole accretion disc is done in Boyer-Lindquist coordinates. The $g$-factor, $g(r)=E_{\rm d}(r)/E_{\rm c}$, is the relativistic shift of the photon energy, $E_{\rm d}(r)$, when emitted from the accretion disc from radius $r$, to the photon energy, $E_{\rm c}$, when received by the corona. The lensing factor, ${\rm d}\Omega_{\rm c}/{\rm d}S(r)$, gives the ratio of the solid angle measured locally in the reference frame of the corona to the Boyer-Lindquist coordinate area on the disc at radius $r$ from where photons are emitted into this solid angle, and we evaluate it numerically. The constants $\sigma$ and $\sigma_{\rm p}$ are the Stefan-Boltzmann constant and its equivalent for the photon density flux.

\subsection{X-ray absorption by the disc}
\label{sec:2p4}

X-rays from the corona illuminate the disc and part of them is absorbed, acting as a second source of energy for the disc. In this case, the disc temperature, $T_{\rm d}(r)$, is set by the energy released by the accretion process locally and by the X--ray absorbed flux as,
\begin{equation}
T_{\rm d}(r) = f_{\rm col}(r)\,\left[\frac{F_{\rm acc}(r) + 2F_{\rm abs}(r)}{\sigma}\right]^{1/4}\ .
\label{eq:temp}
\end{equation}

\noindent $F_{\rm acc}(r)=F_{\rm NT}(r)=\sigma T^4_{\rm NT}(r)$ is the original black-body flux emitted by the disc, with the temperature radial profile given by NT73, and $F_{\rm abs} (r)$ is the absorbed X-ray flux. The factor 2 accounts for the fact that there are two coron\ae, on  either side of the disc. We observe one of them, but the disc is illuminated, and heated, by both of them\footnote{We note that this is another difference with respect to K21a, who did not consider the factor of 2 in their eq.~(2).}. In the case when $L=L_{\rm transf,c}$, then $F_{\rm acc}(r)=0$ at $r\leq$\rtr\footnote{Note that in case of $L=L_{\rm ext,c}$ there is no transition region, i.e., $\rtr=\rms$.}.  This is because all the power dissipated to the disc below \rtr\, is transferred to the corona. All the energy that heats the disc at small radii is due to the X-ray absorption. On the other hand, the temperature above the transition radius is higher than in the NT73 disc, due to the thermalisation of the additional absorbed flux.

Regarding the X-ray flux that is absorbed at each radius (due to disc illumination by the corona located on either side of the disc),

\begin{equation}
    F_{\rm abs} (r) = F_{\rm inc}(r) - F_{\rm refl}(r),
    \label{eq:fabs}
\end{equation}

\noindent where $F_{\rm inc}(r)$ and  $F_{\rm refl}(r)$ are the incident and reflected fluxes, respectively. The incident energy flux, integrated in energy and evaluated per unit local area on the disc including all relativistic effects is  \citep[see e.g.,][]{Dovciak2004ragtime,Dovciak2011}, 

\begin{equation}
    F_{\rm inc} (r) = \frac{g(r)}{U^t_{\!{\rm c}}}\,\frac{{\rm d}\Omega_{\rm c}}{{\rm d}S}(r)\,\frac{L_{\rm X, c}}{4\pi}\ .
    \label{eq:finc}
\end{equation}

\noindent The $g-$factor and d$\Omega_{\rm c}/{\rm d S}(r)$ are the same as in eqs.~(\ref{eq:Lbb}) and (\ref{eq:Nbb}). Like in K21a, for the total reflected flux at each radius, $F_{\rm refl}(r)$, we assume that the local re-processing is given by the {\tt XILLVER} tables \citep{Garcia2013, Garcia16}, which we integrate in energy and over all emission angles. We further assume a radially constant disc density of $n_{\rm H}=10^{15}\,{\rm cm}^{-3}$, an iron abundance $A=1$, and the high energy cut-off value in the disc rest frame (i.e., shifted from the intrinsic value in the corona rest frame to $g(r)\,E_{\rm cut,c}$). 

The ionisation parameter is computed from the incident flux and the disc density. Since the {\tt XILLVER} tables we used are computed assuming an X--ray spectrum with a low energy cut-off fixed at 0.1~keV, we use only the part above $g(r)\,E_{\rm 0,c}$ (i.e., accounting for the relativistic energy shift between the corona and the disc at radius $r$), if $g(r)\,E_{\rm 0,c}>0.1$~keV. This is a small effect in the case of AGN where the seed photon energy is quite smaller than in galactic X--ray binaries, where the seed photon energy can easily be larger than 0.1~keV (see K21a for details).

\subsection{The corona--disc interaction}
\label{sec:discxinteraction}

As we explained in \S\ref{sec:2p2} and \ref{sec:2p3}, we must know the amount of the disc thermal emission entering into the corona in order to compute its X-ray luminosity and size. At the same time, the radiation emitted by the accretion disc depends on its illumination by the corona through thermalisation of the absorbed part of the corona emission (see \S\ref{sec:2p4}), hence on the X--ray luminosity and size.  

To dissect this infinite loop we use an iterative scheme where we initially assume that 
$L_{\rm X,c}=L$, i.e., the second term for the incoming thermal flux in the eq.~(\ref{eq:Lx0}) is neglected. X--rays illuminate the disc, and we compute the disc illumination, eq.~(\ref{eq:finc}), and the absorbed flux, eq.~(\ref{eq:fabs}), that determines the new disc temperature according to eq.~(\ref{eq:temp}). This new temperature profile is then used to compute the incoming thermal disc luminosity and photon number density flux at the corona (eqs.~(\ref{eq:Lbb}) and (\ref{eq:Nbb}), respectively), and to evaluate the average seed photon energy, $E_{\rm 0,c}$, and the average energy of comptonised photons, $E_{\rm sc,c}$, according to eqs.~(\ref{eq:E0}) and (\ref{eq:Ec}), respectively. Finally, these are used to compute the new value of the X-ray luminosity, $L_{\rm X, c}$, according to eq.~(\ref{eq:Lx}) which is then used for the next iteration. If the neglected second term for the incoming thermal flux in the eq.~(\ref{eq:Lx0}) contributes less than $\sim 20-30$\% of $L_{\rm X,c}$ (which is the case for most of the parameter values relevant to AGN),  the iterative scheme works relatively fast. In most cases, after the 6th iteration the values of both $L_{\rm X, c}$ and $E_{\rm 0,c}$ change by less than 1\%, at which point the iterations stop. 

The resulting radial temperature profile, $T_{\rm d}(r)$, $E_{\rm 0,c}$, $L_{\rm X, c}$ and $N_{\rm BB, c}$ are then used to compute the corona radius, the final thermal flux from the accretion disc, and the primary X--ray flux (together with the disc X--ray reflection spectrum), as observed by a distant observer at infinity, for a given set of the basic model parameters (i.e.,  \mbh, \mdot, $h$, and inclination). Note that we include all relativistic effects as detected by an observer at infinity, both from the accretion disc and the X-ray source located on the system axis above the BH \citep[see e.g.,][]{Dovciak2004ragtime,Dovciak2011}.

\section{The broadband SED of the central engine in AGN}
\label{sec:sed}

\begin{figure*}
\centering
\includegraphics[width=0.95\linewidth]{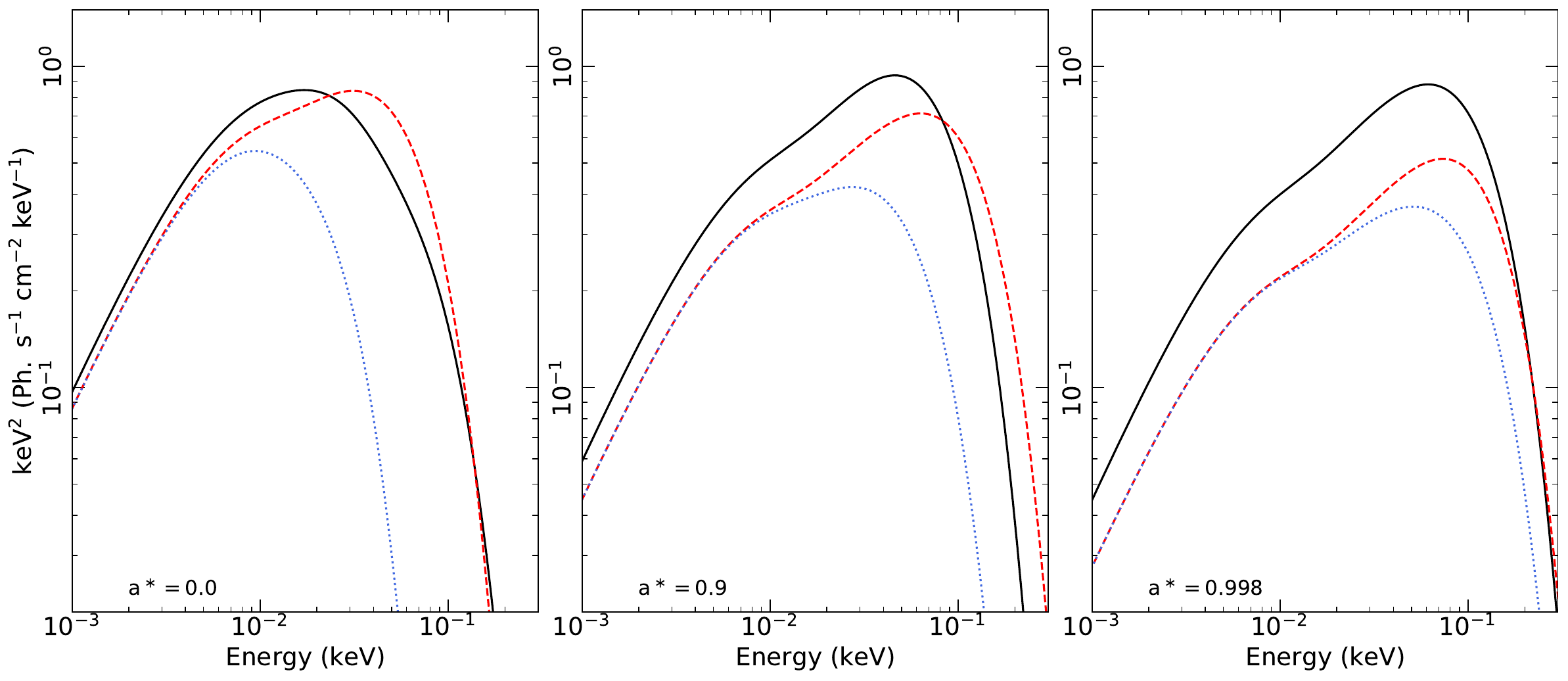}
\caption{NT disc spectra (red, dashed lines), and \kynsed\, disc SEDs with and without thermalisation (black solid lines and blue dotted lines, respectively), for \spin=0, 0.9 and 0.998 (left, middle and right panel, respectively). Physical parameters are: \mbh$=5\times 10^{7}$\msun, \Ltransf~$= 0.5$,  \mdot~$=0.1$, \incl= 40$^\circ$, $h=10\,\rg$, $\Gamma = 2$ and $E_{\rm cut, obs} = 300$~keV.}
\label{fig:thermalisation}
\end{figure*}


An important property of \kynsed\ is that it computes the broadband
SED of AGN, taking into account X--ray illumination of the disc in the lamp-post geometry, assuming that the power that heats the corona is equal to the power the accretion process would dissipate to the gas below \rtr. We discuss below the effects to the SED shape of the energy transfer from the disc to the corona and of the X--ray thermalisation of the disc. We also discuss how the broadband SEDs depend on the main physical parameters of the model.

\subsection{Effects of the energy transfer to the corona and X--ray illumination}

Figure~\ref{fig:thermalisation} shows the effects of energy transfer to the corona and of the X-ray thermalisation on the disc emission in the optical/UV bands. The red dashed lines in all panels show the SED of a NT73 disc (\ntsed), for three different spins. Relativistic effects and colour correction factors are taken into account. The blue dotted line shows the disc SED when half of the accretion power (i.e., all the accretion power released below \rtr$\sim 5, \sim 10$ and $\sim 30$~\rg\, for \spin$=0.998,0.9$ and 0, respectively) is transferred to the X--ray corona and the X--rays do not illuminate the disc, i.e., without thermalisation taken into account. In this case,  both $F_{\rm acc}(r)$ and $F_{\rm abs}(r)$ in eq.\, (\ref{eq:temp}) are zero at $r<$\rtr. Consequently, the disc does not emit below \rtr\,, most of the far-UV emission is missing, and the whole SED is shifted to lower energies. At radii larger than \rtr, the disc emits as a NT73 disc, the disc temperature is the same in both cases, hence both the dashed and dotted SEDs are identical above a certain wavelength.   

The black solid lines show the \kynsed\, SED when X-ray thermalisation takes place. The disc temperature at radii larger than \rtr\, is higher than the NT73 temperature
in this case. As a result, X-ray illuminated discs emit more than the NT73 disc at lower energies (i.e., at longer wavelengths that are produced mainly at larger radii). The effect is more pronounced in the case of high spins because the intrinsic NT73 temperature is lower in these discs at large radii: due to enhanced efficiency, the accretion rate in physical units decreases (for the same accretion rate in Eddington units), and so does the original NT73 disc temperature. Therefore, for the same amount of transferred energy, $L_{\rm transf}$, the effects of X--ray thermalisation increase with increasing spin (see also K21a) and are visible up to higher energies (since \rtr\ is lower). 


At the same time, the disc below \rtr\, is not cold anymore. The disc emits all the way down to \risco, since it absorbs the incident X--rays. This effect is much more pronounced
for small BH spin, where the transition radius \rtr, is much larger for the same amount of the transferred energy, $L_{\rm transf}$.

One can see that the overall SED approaches \ntsed\, in case of a small spin case, since the temperatures in the large region below \rtr ($\rtr \sim 30\,\rg$ for the spin \spin=0) are just redistributed --- the transferred energy, $L_{\rm transf}$, that is used for the illumination, absorption and, eventually, for the thermalisation mainly in this inner region, compensate the original missing NT73 disc emission. Note, however, that for large spins, the transferred energy, $L_{\rm transf}$, is extracted from much smaller region ($\rtr \sim 3\,\rg$ for the spin \spin=0.998) while it illuminates larger region, i.e., increases the temperatures of the original NT73 disc above \rtr\, (compare the black and red lines in all panels of Figure~\ref{fig:thermalisation}). This behaviour will depend on the height of the corona as well as on the amount of the extracted energy (and thus on the size of the \rtr). 

\subsection{Effects of the model parameters to the SED shape}

Figures~\ref{fig:spin} - \ref{fig:mdot} in Appendix show the broadband, \kynsed\, SEDs for various spins, BH masses and accretion rates. In general, the disc temperature increases with increasing spin, decreasing BH mass and increasing accretion rate, as expected. The sharp flux increase at the far-UV, which appears in many SEDs, indicates the low-energy cut-off of the X--ray spectrum, $E_0$. At energies above $E_0$, the disc SED is enhanced due to the contribution from the X--ray continuum.  For the model parameters we have assumed, the model predicts $E_0$ values around $\sim 0.01$ keV (i.e., $\sim 1250$~\AA) when $\mbh=10^9\rm ~M_{\odot}$ (see the upper SEDs in Figure~\ref{fig:mass}), and  when \mdot$\leq 0.1$ (see the respective SED in Figure~\ref{fig:mdot}). The seed photon energy, $E_0$, increases and the flux jump is more pronounced for large spins, because of the higher disc temperature in the inner accretion disc. Due to larger $E_0$ and the same extracted energy, $L_{\rm transf}$, the normalisation of the primary X-ray flux, $A_{\rm c}$, is higher. The flux increase could be observed in the far-UV spectra, for the right combination of parameters, but it will not be easily visible. Most of the times the flux jump is of a low amplitude and, in any case, the flux increase should be smoother than shown in these plots. In reality, we do not expect the low-energy cut-off in the X--ray spectrum to be so sharp.

We kept $L_{\rm transf}/L_{\rm disc}$, $\Gamma$, and the high energy cut-off constant in Figures.~\ref{fig:spin}-\ref{fig:mdot}. The normalization of the X--ray spectrum changes because the total available energy increases with increasing BH mass and accretion rate, but the shape of the X--ray spectrum remains roughly the same. However, at low energies, a significant `excess' emission on top of the power-law continuum below $1-2$ keV is evident in some SEDs. See for example the SEDs when \mbh$\leq 5\times 10^7$, for spins \spin$\geq 0.9$, and when \mbh$\leq 10^7$ M$_{\odot}$, for all spins (Figure~\ref{fig:mass}). This excess flux in these cases is mainly due to the disc thermal emission.
\begin{figure*}
\centering
\includegraphics[width=0.99\linewidth]{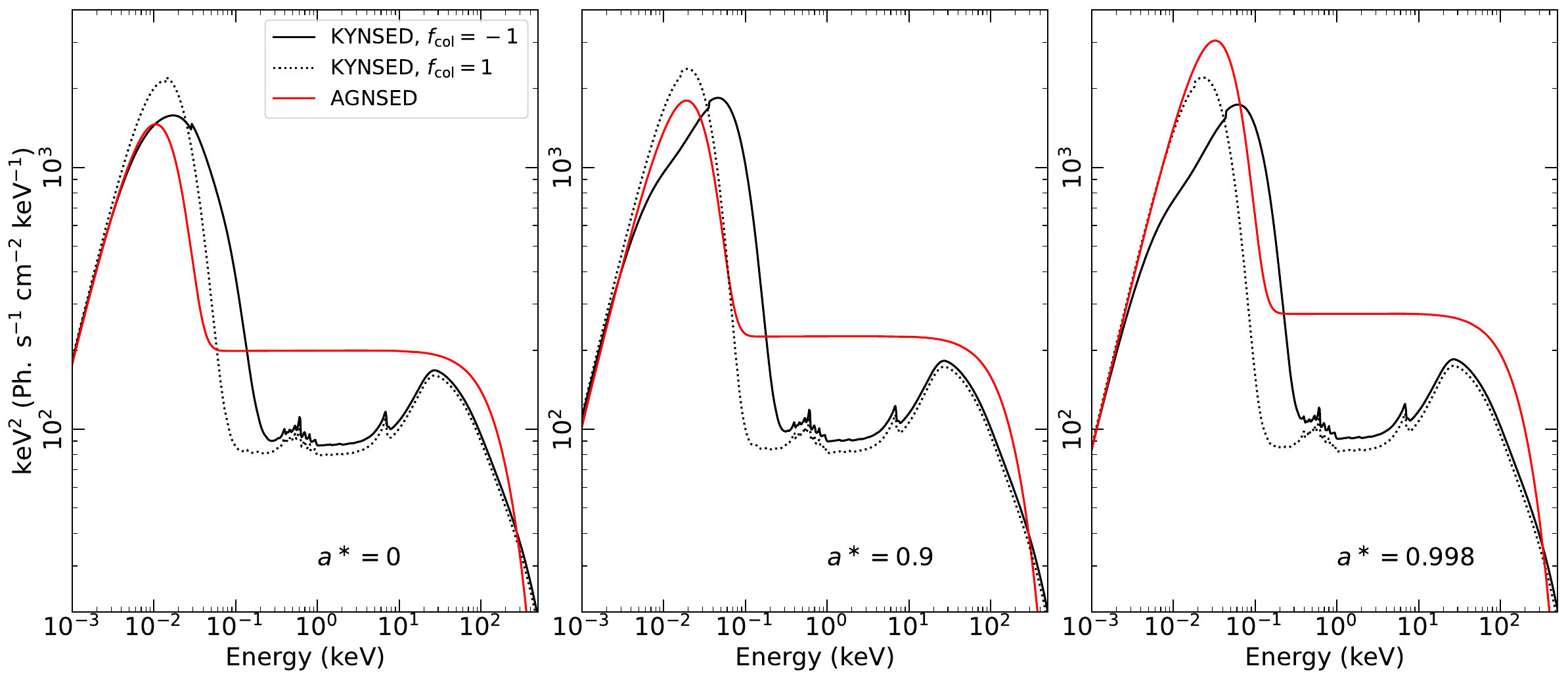}
\caption{Comparison between \kynsed\ and {\tt AGNSED} (red) for $\spin = 0, 0.9,$ and 0.998 (left to right). Inclination is 40 degrees, and $\Gamma=2$ in both models. The black dotted and solid lines show the \kynsed\ model without the spectral hardening effect (colour correction factor $f_{\rm col}=1$) and when the spectral hardening is computed following the \citep{Done2012} prescription ($f_{\rm col}=-1$), respectively.}
\label{fig:agnsed}
\end{figure*}

Figure~\ref{fig:Ltransf} shows the AGN SEDs for various \Ltransf\, values. As expected, the X--ray spectrum normalization increases in proportion to \Ltransf. However, although the fraction of the accretion power transferred to the corona increases, the optical/UV SED does not decrease accordingly. In fact, the optical/UV emission below $\sim 0.05$ keV increases with increasing \Ltransf\, (this shows clearly in the high spin SEDs). This is due to the assumption of isotropic X--ray emission (in the corona's rest frame), which implies that at least 50 per cent of the emitted X-ray luminosity returns back to the disc (more, if the corona height is less than 10 \rg), it gets absorbed, and it increases the disc emission. Even when \Ltransf=0.9, the optical/UV peak is still larger than the SED peak in the X-rays. Note also that two-component thermal emission may be present for very large \Ltransf, especially in the low spin case (see the left panel on Figure~\ref{fig:Ltransf}) --- the two components correspond to the two regions above and below the \rtr. The transition radius is quite large in this case (e.g., $\rtr \sim 200\, \rg$ for spin \spin = 0 and \Ltransf = 0.9, see Figure~\ref{fig:Ltransf_spin}), and the thermal radiation in these two regions may have quite different peak temperatures (especially for low heights when the illumination is concentrated in the region close to the black hole).

The corona height affects the full-band SED significantly. The {\it observed} X-ray luminosity decreases significantly when $h\leq10$ \rg\, (see Figure~\ref{fig:height}), despite the fact that \Ltransf\, remains constant. This is due to light-bending: most of the X--rays fall into the BH or illuminate the inner disc, when the corona is close to the BH. As the corona height increases, the absorbed X--ray flux at larger radii increases, due to an increase in the photon's incident angle (see also K21a). As a result, the disc temperature at larger radii increases in response to the corona height, and so does the flux at longer wavelengths (i.e., below $\sim 0.01$ keV), for all spins. The opposite effect is observed at higher energies. The photon's incident angle in the inner disc does not vary significantly with the corona height. However, as the height increases, the incident X-ray flux on the inner disc decreases, and so does the disc temperature, below \rtr.  Consequently the disc emission above 0.01 keV decreases with increasing height. This is more prominent in the case of large spins.  

The inclination angle affects strongly the disc emission (and the X--ray reflection spectrum), but not the X--ray continuum, since we have assumed that the corona emission is isotropic in its rest frame. At high inclinations ($\theta \geq 70^{\circ}$), the disc flux is substantially decreased (see Figure~\ref{fig:inclination}). The flux jump at $\sim 0.03-0.05$ keV is most evident in this case. If the molecular torus does not block the view to the central nucleus in luminous AGN,  this feature should be observed in the far-UV spectra of highly inclined sources at redshifts larger than $\geq 3$ (for the parameters chosen in Figure~\ref{fig:inclination}).

Figures.~\ref{fig:gamma} and \ref{fig:ecut} show the effects of $\Gamma$ and \ecut\, in the SED. A softer spectrum results in a stronger soft-excess (around 0.3--1 keV), and stronger flux jump at $E_0$ (Figure~\ref{fig:gamma}). The X--ray spectrum normalization at 1 keV decreases with increasing \ecut\, (Figure~\ref{fig:ecut}), despite the fact that \Ltransf\, remains constant. This is because, once \rtr\, is determined, both \Ltransf\, and $E_0$ are fixed (for a given combination of \mbh, \mdot, \spin and $h$). Therefore, the X--ray spectrum normalization is uniquely determined from the photon index, $\Gamma$, and \ecut. For the same $\Gamma$, the normalization decreases as \ecut\, increases, and the (fixed) X--ray luminosity is spread over a larger energy band. 

\subsection{Comparison with AGNSED}

Figure~\ref{fig:agnsed} shows a comparison between \kynsed\, and {\tt AGNSED} for an AGN with \mbh=$5\times 10^7~\rm M_\odot$ and \mdot$=0.1$, for \spin$ = 0, 0.9, 0.998$. The black dotted line shows the \kynsed\ SED when $f_{\rm col}=1$ (which is what {\tt AGNSED} assumes) while the black solid line shows the \kynsed\ model when we adopt the \cite{Done2012} prescription. 

To the best of our knowledge, {\tt AGNSED}  is the only other model for the broadband AGN SEDs (like \kynsed). 
The accretion flow in {\tt AGNSED} is assumed to be radially stratified. It emits as a standard (i.e., NT73) disc blackbody from the outer disc radius to $R_{\rm warm}$, and in the region between $R_{\rm warm}$ and a smaller radius, called $R_{\rm hot}$, the model assumes the passive disc scenario of \cite{Petrucci18}. This is the so-called ``warm Comptonization'' region, which is supposed to be responsible for the observed soft X--ray excess in AGN. At a radius below $R_{\rm hot}$, the disc is truncated. Like \kynsed, {\tt AGNSED} assumes that the accretion power that would be dissipated in the disc below $R_{\rm hot}$ is transferred to the hot Comptonisation component, which is responsible for the X--ray continuum in AGN. 

We have assumed $\Ltransf=0.5$ in \kynsed\, for the spectra shown in Figure~\ref{fig:agnsed}. We also set $R_{\rm hot}=31.55, 10.96,$ and 4.77~\rg\ for the {\tt AGNSED} spectra shown in the same figure, for $\spin = 0, 0.9,$ and 0.998, respectively, so that the energy transferred to the hot Comptonisation component will be equal to half the total accretion power. We set the corona height at 10~\rg \,and \ecut\ at 300~keV in \kynsed. For {\tt AGNSED}, we do not consider the warm corona, and we set $kT_{\rm e}=150$~keV. 

An obvious difference between the two models is the presence of the X--ray reflection features in \kynsed. This is because {\tt AGNSED} does not consider X--ray reflection from the disc. There is a large difference between the X--ray continuum normalization in \kynsed\, and {\tt AGNSED}. To a large extent, this must be due to the fact that the corona in \kynsed\, emits isotropically to all directions, so half of the X--ray power is directed away from the observer to the disc (which extends to \risco\, and blocks emission from the corona which is located on the other side of the disc). On the other hand, there is no disc (physically) below $R_{\rm hot}$ in {\tt AGNSED}, hence the observer can detect emission from the whole of the corona. 
The \kynsed\, and {\tt AGNSED} SEDs show differences in the optical/UV bands as well, at all spins. There are significant differences even when we adopt $f_{\rm col}=1$. In the \spin=0 case, the difference in the far UV should be due to the X--ray heating of the inner disc (below \rtr) in \kynsed. The disc in {\tt AGNSED} does not exist below $R_{\rm hot}$, while the disc extends down to \risco\ in \kynsed. Although the power dissipated due to the accretion process below \rtr\ is transferred to the corona, the disc is illuminated by the X--rays. Most of the them are absorbed, and the disc is heated. As a result, the disc emits, mainly in the UV, hence the difference with {\tt AGNSED} in these wavelengths. 

At higher spins, the {\tt AGNSED} optical/UV flux increases and in fact overcomes the \kynsed\ flux (below $\sim 0.01$ keV) when \spin=0.998, despite the fact that the disc in \kynsed\ emits down to \risco, while the disc is truncated at $R_{\rm hot}$ in {\tt AGNSED} (below which half of the disc power is emitted).
This discrepancy is probably due to the fact that {\tt AGNSED} does not take into account relativistic effects. Due to GR effects a large amount of flux from the inner disc will end up in the BH and the disc, hence the difference in the two spectra. 

The comparison between the black solid and dotted lines in Figure~\ref{fig:agnsed} shows that the choice of $f_{\rm col}$ affects significantly the optical/UV spectrum. For the adopted model parameter values, the \citep{Done2012} prescription significantly alters the spectrum in the optical/UV band. In fact, the choice of $f_{\rm col}$ affects the X--ray spectrum as well, as $E_0$ depends on the disc temperature (although the differences between the solid and dotted lines in the X-ray band are not that significant).

Note that we have assumed $\Ltransf=0.5$ for the KYNSED spectra on Figure~\ref{fig:agnsed}. This implies that the same amount of power that is released as the disc thermal radiation is also transferred to the corona. This power is used to heat the electrons, and then it is assumed to be released, in its entirety, as X-ray luminosity (see eq.~(\ref{eq:Lx0}), which shows that the X-ray luminosity is equal to the total power transferred to the corona plus the luminosity of the soft photons that are scattered in the corona). However, as can be seen e.g., on the left panel of the figure, the total X--ray flux appears to be much less than the total thermal flux. 

One reason for this is that the thermal energy is emitted as the cosine source (i.e., as $\mu\,L/\pi$; here, $\mu=\cos{\theta}$ is the cosine of the inclination angle, and $L$ is the power given to the disc and to the corona, which are assumed to be equal), while the X-ray source is assumed to be isotropic (i.e., $L/4\pi$). A second reason is that half of the X-ray emission is incident to the disc, where most is absorbed (for neutral disc it can be as high as 80\%) and the rest is reflected in X-rays (the other 20\%). Thus, the total thermal radiation should be $\mu\,L\,(1+0.8/2)/\pi$, while the total X-ray emission is $L/4\pi+\mu\,0.2\,L/2/\pi$. This approximate estimate predicts a ratio between the observed X-ray emission and the disc thermal radiation of 30\% (for a $40\degr$ inclination). A precise measurement shows that the total observed X-ray flux makes up only 22\% of the total observed thermal flux in the \kynsed\ spectra shown in the left panel of Figure~\ref{fig:agnsed}. The difference with the result from the approximate computation is due to the fact that in the latter case we exclude, of course, all the nuances of the computations, including radial dependence of the emission and illumination, relativistic effects etc.  

\section{Fitting the average, broadband spectrum of NGC~5548}
\label{sec:fitting}


As an example of how \kynsed\ works in practice, we chose to fit the average SED of NGC~5548. This is a typical Seyfert 1 galaxy, at a distance of $D_{\rm L} = 80.1~\rm Mpc$ according to NED\footnote{The NASA/IPAC Extragalactic Database (NED) is funded by the National Aeronautics and Space Administration and operated by the California Institute of Technology.}. It has been observed extensively in the past, at all wavebands. In particular, NGC~5548 was observed by {\it Swift}, {\it HST}, and many ground based optical telescopes from February to July 2014. These observations were part of the Space Telescope and Optical Reverberation Mapping project \citep[STORM;][]{Derosa15}. {\it Swift} observed the source over 125 days with a mean sampling rate of $\sim 0.5$~day in X--rays and across six UV/optical bands. Approximately daily {\it HST} UV observations were also obtained. The ground-based optical monitoring observations resulted in  light curves which have nearly daily cadence in nine filters, namely {\it B,V,R,I} and {\it u,g,r,i,z} \citep[see e.g.,][]{Edelson15,Fausnaugh16}. These observations resulted in X-ray/UV/optical light curves which are among the densest and most extended light curves ever obtained for an AGN. \cite{Edelson15} and \cite{Fausnaugh16} used these light curves to compute the X--ray to UV/optical time-lags, and K21b fitted them using the K21a model. Therefore, it is interesting to fit the  average SED of NGC~5548 using the same data set that K21b used to fit the optical/UV time-lags, and investigate whether the same model can fit both the average SED and the X--ray/optical/UV time-lags with the same best-fit parameters. K21b fitted the time-lags of other AGN as well, but we chose NGC~5548  because the host galaxy contribution in the optical/UV bands is quite well known for this object \citep[see][]{Mehdipour15,Fausnaugh16}.

\subsection{The average SED}
We used the data from the 2014 multiwavelength campaign to compute the mean SED of NGC~5548.
Regarding the UV/optical band, we used the data in \cite{Fausnaugh16}. These authors computed the mean of all light curves in 18 spectral bands (from $\sim 1160-9160$~\AA) and list the results in Column~8 of their Table 5. Mean fluxes are corrected for Galactic extinction, and their error is equal to the rms scatter of the points in the light curve. \cite{Fausnaugh16} also list the host galaxy contribution in each band, which we subtracted from the mean observed flux. 

As for the X-rays, first we considered the $2-10$~keV {\it Swift}/XRT light curve, and we noticed that the count rate in the time interval between $56712-56714$~MJD, $56775-56785$~MJD, and $56807-56812$~MJD is close to the overall mean count rate. There are 5, 20, and 14 observations during these intervals, respectively. We used data from all the observations in each interval and we extracted the overall X–ray spectrum of each interval using the automatic {\it Swift}/XRT data products generator\footnote{\url{https://www.swift.ac.uk/user_objects/}} \citep{Evans09}. We considered the $0.4-8$ keV band in each spectrum, and grouped the spectra to have at least 25~counts per bin.

The flux measurements from broadband UV and optical filters can be affected by emission from the Balmer continuum, blended Fe {\small II} and other emission lines that originate in the broad line region \citep[BLR; e.g.,][]{Korista01, Korista19}. \cite{Fausnaugh16} estimated that the Balmer continuum emission accounts for about 19\% of the flux in the $u$  and $U$ filters, and the H$\alpha$ line contributes $\sim 20$ and 15\% of the flux in the $r$ and $R$ bands, respectively. We therefore excluded these points when we fitted the data. We also excluded the {\it Swift}/UVOT M2 and W1 data (in the $2000-3000$~\AA\ range), as they may also be affected by the Balmer continuum emission. 

\begin{figure}
\centering
\includegraphics[width=0.99\linewidth]{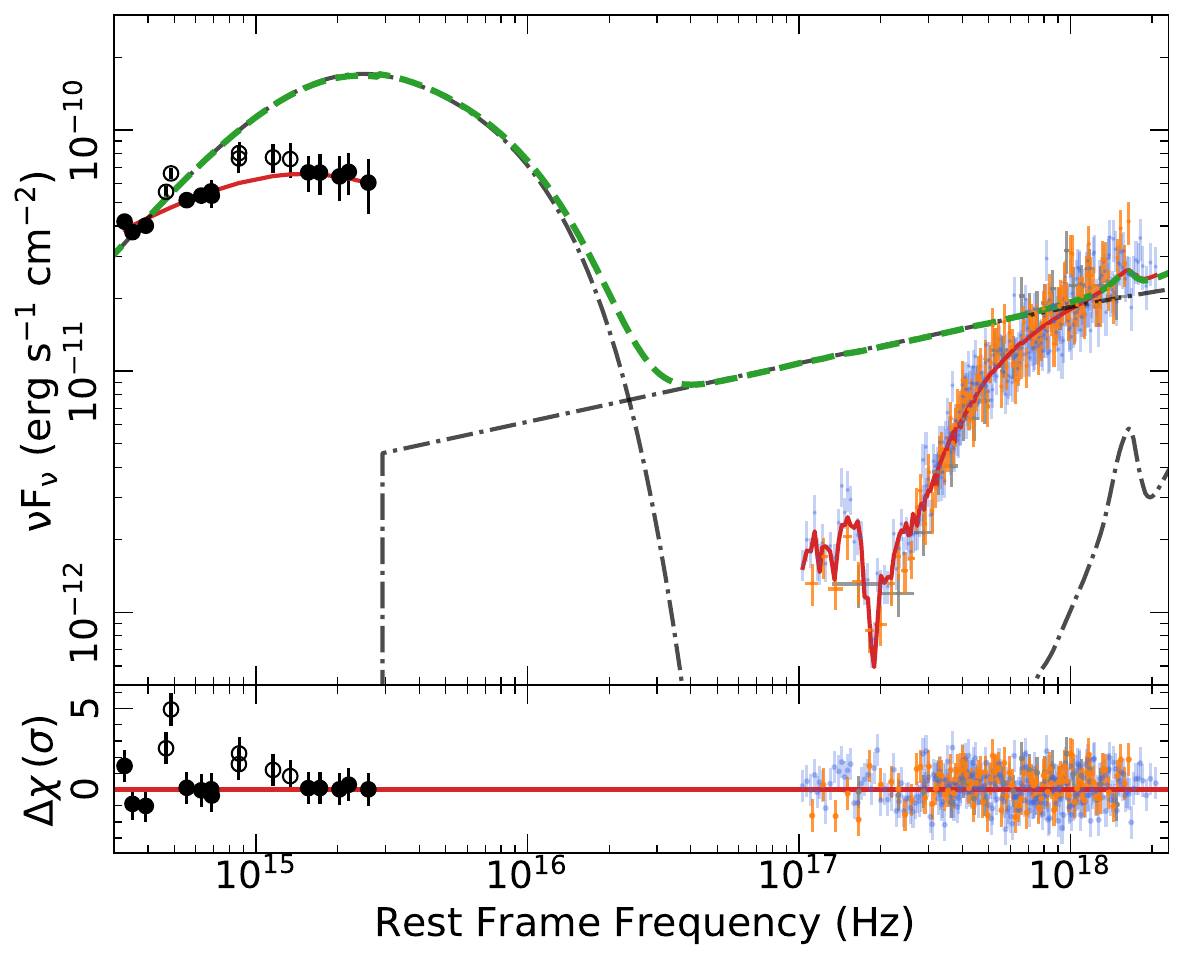}
\caption{The observed broadband NGC~5548 SED. The solid red line shows the model obtained for $h = 46$~\rg. The green dashed line shows the unabsorbed best-fit model (i.e., by removing all Galactic and intrinsic, neutral and warm, absorbers). The black dash-dot lines show the various spectral components. Empty circles indicate the data points that we did not consider during fitting, to avoid the contamination from the BLR (see Section~\ref{sec:fitting} for details). The grey, blue, and orange data points indicate the three X-ray spectra we used in the fitting.}
\label{fig:ngc5548spectrum}
\end{figure}

\subsection{The theoretical model}
We fitted the three X-ray spectra and the UV/optical data simultaneously. The spectral fitting is performed using XSPEC \citep{Arnaud1996}. We used the FTOOLS command {\tt ftflx2xsp}\footnote{\url{https://www.heasarc.gsfc.nasa.gov/lheasoft/ftools/headas/ftflx2xsp.html}} to create spectral files in the UV/optical range in a format that is compatible with XSPEC. The overall model, written in XSPEC parlance, is  as follows,
\begin{eqnarray}
\rm Model &=& {\rm \tt extinction_{Czerny} \times \kynsed_{UV/opt.} } \nonumber \\
& + & {\rm \tt TBabs \times zTBabs \times zxipcf  \times \kynsed_{\rm X-rays}}.
\label{xspecmdel}
\end{eqnarray}

\noindent We use \kynsed\, twice; one for the optical/UV and the other for the X-ray spectra. 
Obviously, all the relevant \kynsed\, parameters were tied in the UV/optical and X-rays range. The {\tt TBabs} component \citep{wilms00} accounts for the effects of the Galactic absorption in the X--ray band (the UV/optical data points are already corrected for Galactic extinction). We fixed its column density to $1.55\times 10^{20}~\rm cm^{-2}$ \citep{HI4PI}\footnote{All model parameters were tied together in the X-ray spectra, as they are all considered to be representative of the mean X--ray spectrum, i.e., they are treated as data sets of the same intrinsic spectrum.}. 

\begin{figure}
\centering
\includegraphics[width=0.98\linewidth]{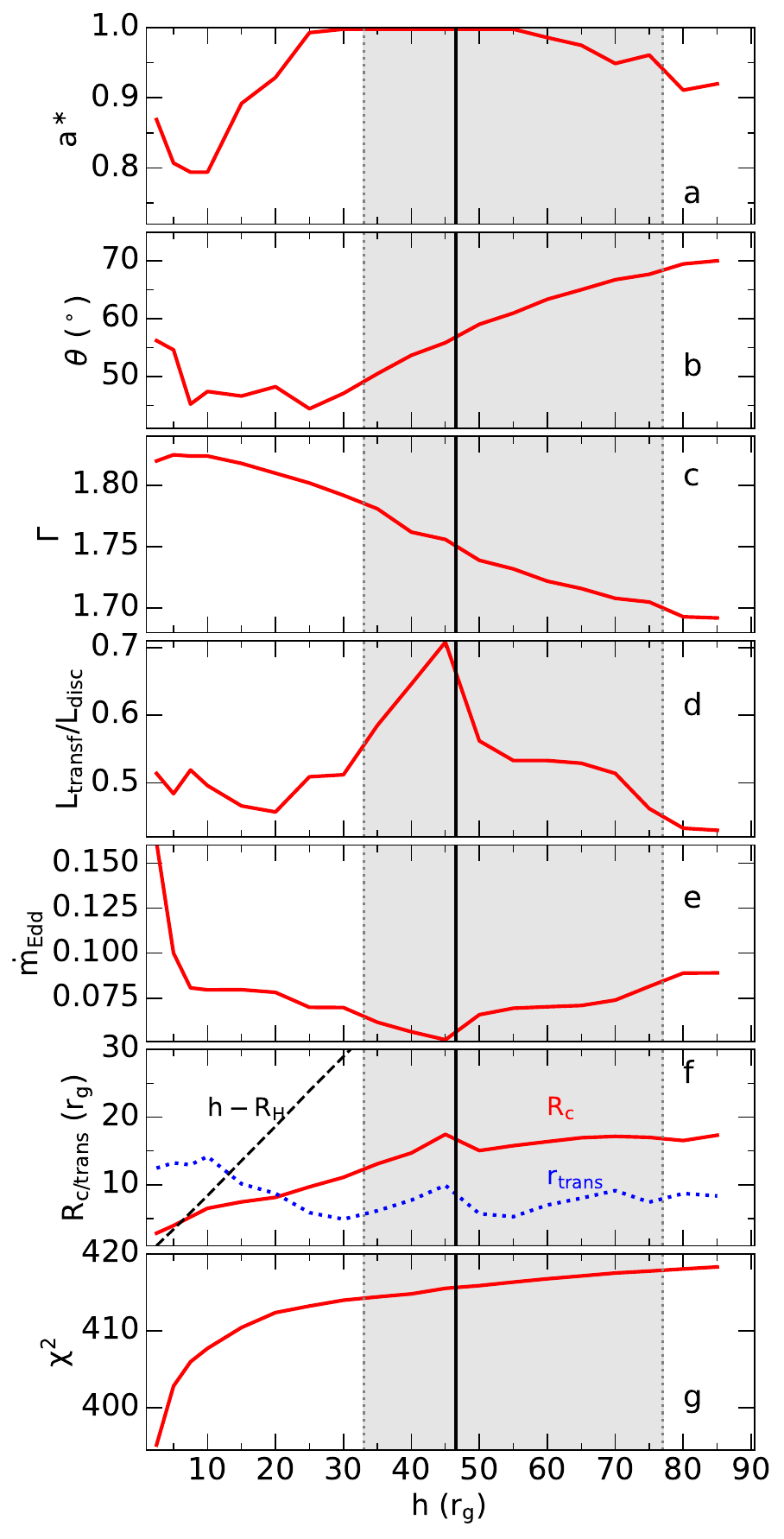}
\caption{The best-fit parameters obtained by fitting the NGC~5548 broadband spectra for various corona heights. The vertical solid line and the grey shaded area indicate the best-fit value and the $1-\sigma$ confidence range, respectively, of the height obtained by fitting the time-lag spectrum of this source with $a^\ast = 1$ in K21b. The black dashed line in panel $f$ shows the difference between the corona height and the event horizon, and the blue dotted line in the same panel shows the best-fit \rtr\, that corresponds to the fitted values of transferred energy, \Ltransf, and black-hole spin, \spin.}
\label{fig:par}%
\end{figure}

\subsection{Host galaxy absorption}
We also considered the possibility of intrinsic absorption in the host galaxy. We considered the extinction curve of \cite{Czerny04} for the absorption in the optical/UV. The {\tt extinction$_{\rm Czerny}$} component is determined by $E(B-V)_{\rm host}$, which we left free during the model fitting. The {\tt zTBabs} component accounts for the absorption in the X--ray band of the same absorber, and its column density was kept linked to $E(B-V)_{\rm host}$  through the relation $N_{\rm H, host} =  5.8 \times 10^{21}~E(B-V)_{\rm host} ~ {\rm cm^{-2}}$ \citep{Bohlin78}. Finally, we used the spectral component {\tt zxipcf} \citep{Reeves2008}, to account for possible warm absorption. This component accounts for absorption from partially ionised absorbing material, which covers some fraction of the source, while the remaining of the spectrum is seen directly. We left all the parameters of this component free during the model fitting. 

\subsection{The fitting process}
We chose to fit the data by assuming that the X-ray source is powered by the accretion process. In addition to \rtr\, the other \kynsed\, physical parameters in this case are black hole mass and spin, \Ltransf, accretion rate, corona height, photon index, high energy cut-off, and inclination. We fixed the BH mass at $5.2\times 10^{7}~\rm M_\odot$ \citep{Bentz15}, and we assumed a high energy cut-off of 150~keV \citep{Ursini15}. Still, there are too many parameters to constrain by fitting the data. For example, the figures in the Appendix show that spin, accretion rate and inclination affect significantly the optical/UV SED. Actually, the corona height also affects the optical/UV, as well as the X--ray normalization, together with \Ltransf\, and $\Gamma$. We therefore decided to fit the data by considering a grid of height values,  from $h=2.5$~\rg\ to $h=10$~\rg, with a step of 2.5~\rg, and from 10 to 80~\rg\, with a step of 5~\rg. At each height, we left the spin, inclination, photon index, \Ltransf, and \mdot\ free to vary. 

\subsection{The best-fit results}
The model fits the data very well, at all heights,  with $p_{\rm null} > 0.1$. We believe this is due to the effect we mention in previous subsection: many model parameters affect the shape of the SED. As height changes, other physical parameters like, e.g.,  \mdot, \Ltransf, and inclination, also change in such a way so that the final fit is always acceptable. The best-fit values of each of the parameters are shown as a function of height in Figure~\ref{fig:par}. The plots show that spin varies from $\sim 0.8-1$, \Ltransf\ from $\sim 0.4-0.7$, and \mdot\ from $\sim 0.05$ to 0.15 (a factor of 3 difference), as we increase the height from $\sim 5$ to 85 \rg.

The best-fit parameters for all absorption components do not depend on the corona height (that is why we do not plot them in Fig.\,\ref{fig:height}). We found $E(B-V)_{\rm host} = 0.12 \pm 0.02~\rm mag$ for the host galaxy extinction (all errors are $1\sigma$ errors, unless otherwise stated). This is consistent with the $E(B-V)$ value of 0.09$^{+0.08}_{-0.07}$ that \cite{Panagiotou20} reported from the modelling of the optical/UV PSDs (which were estimated using light curves from the 2014 monitoring campaign as well). \cite{Kraemer98} reported a reddening of $E(B-V)=0.07^{+0.09}_{-0.06}$ of the narrow line region in NGC~5548. Since the portion of the reddening that is due to our own Galaxy is small \citep[$\sim 0.016$,]{Schlafly2011}, this value is also consistent with our measurement. In addition, in our modelling, the absorber in the optical/UV band is linked to absorption by neutral gas in the X-ray band. The best-fit $E(B-V)$ value implies a neutral absorber with N$_{\rm H}\sim 6\times 10^{20}$ cm$^{-2}$. This is almost identical to the N$_{\rm H}$ measurement of \cite{Mathur17}, when they fit {\it Chandra} spectra taken during the STORM campaign, with a power law plus a warm corona model.  As for the ionised absorber, we found that $N_{\rm H} \sim 2 \times 10^{22}$ cm$^{-2}$, $\log (\xi / \rm erg~cm~s^{-1}) \sim 1$, and $f_{\rm cov} = 0.94-1$. 

The fit results in a high spin value of \spin$> 0.8$, for all heights (see top panel in Fig.\,\ref{fig:par}). This is in agreement with the results obtained by K21b, who found that the $\spin=1$ case was preferred, because the resulting accretion rate was consistent with the value of $\mdot=0.05$ that is reported in the literature for NGC~5548. 

As we have already discussed in \S\ref{sec:2p2}, \kynsed\, also computes $R_{\rm c}$, assuming conservation of the photon number during Comptonization. This model parameter is not used for fitting the data, but it is computed when model fitting is finished, and can be viewed using the {\tt XSPEC} command {\tt xset}. The corona radius can be used to check the physical consistency of the model, since $R_{\rm c}$ should be smaller than the height of the X--ray source minus the event horizon radius, $R_{\rm H}$. The black dashed line in panel (f) of Figure~\ref{fig:par} shows $(h - R_{\rm H})$ as a function of $h$. Below $h = 7.5$~\rg, $(h - R_{\rm H})$ is always smaller than $R_c$. Clearly, despite the small $\chi^2$ values, we reject all heights smaller than $h=7.5$~\rg\ as physically unacceptable solutions. 

\begin{table}
\centering
\caption{The range for the main physical parameters in NGC~5548 which corresponds to the $1-\sigma$ limit on height that is obtained for fitting the time-lags in K21b. See \S\ref{sec:fitting} for more details about the estimation of these limits.}
\label{tab:bestfit}
\begin{tabular}{ll} 
\hline
\spin	&	> 0.94	\\
$h~(\rg)$	&	[33, 77]	\\
\mdot	&	[0.052, 0.085]	\\
\Ltransf	&	[0.45, 0.70]	\\
$\Gamma$	&	[1.7, 1.78]	\\
$\theta$	&	[49\degr, 68\degr]	\\
$R_{\rm c}~(\rg)$	&	[12, 17.5]	\\
$r_{\rm trans}~(\rg)$	&	[5, 10]	\\
\hline
\end{tabular}
\end{table}

We can use the results from the K21b fits to the time-lags spectra of NGC~5548 to put additional constrains to coronal height range that would be acceptable from a physical point of view. 
We therefore considered the 1$\sigma$ confidence region for height from fitting the time-lag spectrum in K21b ($33-77$~\rg\ for $a^\ast = 1$), to constrain even further the best-fit results to the mean SED. The respective $1\sigma$ confidence region on the height is indicated by the grey shaded area in Figure~\ref{fig:par}.  The solid vertical line in Figure~\ref{fig:par} indicates the best fit height  we obtained in K21b ($h=46$~\rg). The fit to the energy spectrum in this range is statistically acceptable, and predicts a very high spin value ($a^\ast > 0.94$). In fact, this result indicates how powerful it is to combine spectral and timing information to constrain the model. K21b could fit the time-lags equally well, both when assuming $ \spin=0$ or 1. When combined with the fitting of the SED, the hypothesis of $\spin=0$ is not accepted any longer. 

Table~\ref{tab:bestfit} lists the range of the best-fit model parameter values that correspond to heights in the $33-77$~\rg\ interval. The transition radius range of $\sim 5-10$ \rg\ implies a rather small disc region, close to ISCO, where accretion power is transferred to the corona. However, given the fact that the best-fit BH spin is high, this power is large (at least $\sim $50\% of the total disc luminosity is transferred to the corona). 

The best-fit accretion rate range is fully consistent with the $\sim 5$\% value,  which is based on bolometric luminosity estimations \citep{Fausnaugh16}. It is also consistent with the accretion rate estimates from the fitting of the time-lags spectra. K21b found \mdot=0.03, with an 1$\sigma$ confidence region of [0.005 - 0.05] (see their Table 3). Our 1$\sigma$ confidence region is almost overlapping with theirs, indicating the agreement between the two approaches (within 1$\sigma$). We note that K21b assumed $\Gamma=1.5$, based on the results of \cite{Mathur17}. This is different to our best-fit results of $\Gamma \sim 1.75$, but the time-lags do not depend on $\Gamma$, as long as $\Gamma\leq 2$ (see Fig.~19 in K21a). Therefore, the K21b model would fit the NGC~5548 time-lags equally well even if they had assumed a photon index of 1.7-1.8. 
Our best-fit results of $\sim 50$\degr$-70$\degr for the disc inclination is rather large, for a Type 1 AGN. Interestingly, \cite{Pancoast2014} found that the geometry of the H$\beta$ BLR in NGC~5548 is that of a narrow thick disc (see Fig.\,1 in their paper),
with an inclination angle of 38.8$^{+12.1}_{-11.4}$ degrees. This agrees (within 1$\sigma$) with our results, which implies that the inclination of the BLR disc and the accretion disc may be similar in NGC5448, and that it may be slightly larger than 40 degrees, which is usually assumed for Type I AGN.

As described in the \kynsed\ model documentation, the {\tt XSPEC} command {\tt xset} outputs various computed values of the system properties like, e.g., the value of the ionisation parameter at the inner and outer edge of the accretion disc, the intrinsic and observed X-ray luminosity of the primary X-ray source in the 2-10 keV energy band, the reflection ratio, the ISCO and transition radius, as well as the optical depth and the electron density in the corona. For height 46\rg\ (indicated by the vertical black line in Fig.\,\ref{fig:par}), we obtain the best-fit to the model for $a=0.998$, $\theta=55.9^\circ$, $\Gamma=1.756$, $\Ltransf=0.68$, and $\mdot=0.052$ (see Fig.\,\ref{fig:par}). Then, the command {\tt xset} lists 
the corona radius, $R_{\rm c}=15.3\,\rg$, optical depth, $\tau=1.2$, electron column density, $\Sigma_{\rm e}=\tau/\sigma_{\rm T}=1.8\times10^{24} {\rm cm}^{-2}$ (where $\sigma_{\rm T}=6.652\times10^{-25}{\rm cm}^2$ is Thomson cross-section), and the electron density in the local frame of the corona, $n_{\rm e, c}=\Sigma_{\rm e}/(2R_{\rm c}\sqrt{U_{\rm c}^t})=7.8\times10^9 {\rm cm}^{-3}$.

Figure~\ref{fig:ngc5548spectrum} shows the mean optical/UV and X--ray SED of NGC~5548, constructed as we explained above. The solid and the dashed lines in the same figure correspond to the absorbed and unabsorbed best-fit model, respectively, when $h = 46$~\rg. The dot-dashed lines in the same figure show the individual spectral components of the best-fit  model to the broad-band SED, i.e. the disc thermal emission (including thermalization due to the absorbed X-rays), the X--ray emission from the corona, and the X--ray reflection component (both the overall and the individual spectral components are plotted as seen by a distant observer). The model fits the data well ($\chi^2=415.6/388$~dof). Open symbols in Figure~\ref{fig:ngc5548spectrum} show the $M2, W1, U,u, R$ and $r$ band measurements that we did not consider during the model fitting. The observed average flux in those bands is indeed larger than the best-fit model. We estimate this excess to be $\sim 10-30$~\% in the $M2$ to $U$ bands, and $\sim 20-40$~\% in the $r-$ and $R-$bands, respectively. Those values are consistent with the ones estimated by \cite{Fausnaugh16} as the contribution of the BLR to the observed flux in the same bands.

\section{Discussion and conclusions}
\label{sec:discussion}

We present a new model for the broadband spectral energy distribution, from optical to X--rays, of AGN. The model includes an X-ray illuminated standard NT73 disc, where all the relativistic effects are taken into account. We also include a colour correction factor, following the prescription of \cite{Done2012}. The model assumes the X-ray corona is located at height $h$ above the BH (the lamp-post geometry), and emits isotropically in its rest-frame. We take into account all the relativistic effects when calculating the photons path from the disc to the corona and back, and from the disc and the corona to the observer at infinity. 

The \kynsed\ model can be used in the {\tt XSPEC} spectral fitting package. It is publicly available, and it can be downloaded from {\tt https://projects.asu.cas.cz/dovciak/kynsed}. A file called KYNSED-param.txt describes all the model parameters and can be downloaded from the same web address.

\subsection{The X--ray luminosity}
The power that is given to the corona can be either a free parameter or it can be linked to the accretion power. In the second case, 
the model assumes that the total accretion power that is dissipated below a transition radius, \rtr, is transferred to the X--ray corona. The mechanism that feeds the corona may not be 100\% efficient, and the power that is transferred to the hot corona may not be exactly equal to $L_{\rm transf}$, as defined by eq.\, (\ref{eq:ltransf}). In fact, the fraction of the power dissipated to the disc when the cold matter accretes over the power which is transported to the corona could be an extra free parameter of the model, but this is not currently incorporated in the model. In its current version the model with a particular transition radius, \rtr, gives an upper limit of the X-ray normalisation.

It is also possible that the corona is powered by another source of energy, for example by the Blandford-Znajek process \citep{Blandford77}. This mechanism is assumed to power the jets in radio-loud AGN. This possibility can be accommodated in the present model, if \rtr\ is set equal to \rms. In this case, 
the energy that is provided to the corona, $L_{\rm ext}$, is a free model parameter. The only difference then is that the disc emits, like a NT disc, down to \risco, and it is also heated by the X-rays which take the power from somewhere else.

\subsection{The low-energy X--ray cut-off}

We  take  into  account  all  transformations  to  the  photons energy (due to GR) from the disc to/from the corona rest-frames, and to the observer. In particular, \kynsed\ computes self-consistently the low-energy cut-off in the X–ray spectrum, $E_0$, using  an  iterative  process,  which  takes  into  account  the  average energy of the disc photons arriving to the corona (in its rest frame). This depends on the accretion rate and \rtr, but also on  the X–ray flux illuminating the disc (which, in turn, depends on $E_0$). 

The correct estimation of $E_0$ is important for two reasons. On the one hand, if the spectral slope and the high energy cut-off can be determined by the X–ray data, then the X-ray normalization is uniquely determined, as long as $E_0$ is set. This may have important implications in the determination of the X–ray continuum amplitude and the study of the iron-line shape in the X–ray  spectra  of  AGN.  Secondly, $E_0$ is  necessary  for the  accurate calculation of the incident, reflected and absorbed X–ray flux on each disc element above and below \rtr\, and hence on the determination of the disc heating and its temperature. 

\subsection{The corona height and size}
\label{sec:discussion-corona-size}

\kynsed\ may overestimate  the corona height. The model assumes a flat disc, which is not in agreement  with  the  standard  SS73  and  NT73  predictions.  Although small, the disc height over radius ratio in these models is not zero. We expect the disc height to increase with increasing radius. Even a small increase of the disc height will affect the incident angle of the X-ray photons arriving on the disc surface. This will increase the incident, and the absorbed X–ray flux in each disc element. It is therefore possible that coron\ae\ with smaller heights will be able to heat the disc more efficient, and perhaps fit the data with the same quality when the height is larger. We plan to update \kynsed\ with an advanced treatment of the X–ray illumination of discs with non-zero height in the future.

\begin{figure}[t]
\centering
\includegraphics[width=\linewidth]{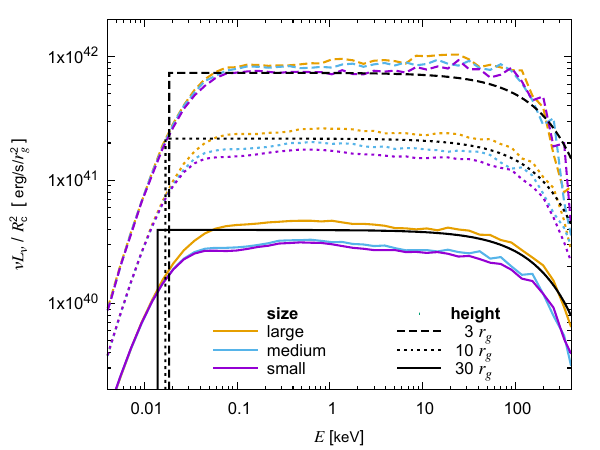}
\caption{Comparison of the primary X-ray spectra in the point source approximation used in \kynsed\/ (black lines) with the 3D corona Monte Carlo computations of {\tt MONK} (colour lines) for the spherical corona with different size, small (magenta), medium (blue) and large (yellow), at three heights. The three shown radii of the corona are: $R_{\rm c}=0.5, 1.5$ and $2\,\rg$ for the height $h=3\rg$; $R_{\rm c}=1, 6$ and $9\,\rg$ for the height $h=10\rg$; and $R_{\rm c}=5, 15$ and $29\,\rg$ for the height $h=30\rg$. Note that the corona with the largest size extends as far as the BH horizon. The luminosities are divided by $R_{\rm c}^2$ so that we normalise the results to the same (unit) radius. See Section~\ref{sec:discussion-corona-size} for more details.}
\label{fig:MONK-KYNSED}
\end{figure}

The model computes the X--ray corona size using the conservation of photons during the Comptonisation process. This computation is approximate, as it assumes a fixed corona temperature, and an approximate relation between the corona optical depth and its temperature (see \S \ref{sec:2p2}). Nevertheless, an inspection of the corona size after fitting the data can provide useful constrains on the validity of the model. If the radius is considerably larger than its height (say by a factor of $\sim 2$, or more) this would indicate that the adopted lamp-post geometry is not consistent with the observed SED \cite[see also][where the estimation of the corona size from 3D computations is discussed]{Ursini20}.

To check the validity of the point-source approximation, we have computed the Comptonised spectra emerging from the corona with realistic (non-infinitesimal) size with the 3D Monte Carlo code {\tt MONK}, see \cite{Zhang2019}, which includes all relativistic effects. We considered three heights of the X--ray corona, $h= 3, 10$ and $30 \rg$, and  
we considered several sizes, from a small one to a very large corona that would actually reach as far as the BH horizon. The corona was assumed to be static and homogeneous with an optical depth that would result in a power-law like energy spectrum with an index $\Gamma\approx 2$. The electron temperature was set to $T_{\rm e}=100\,$keV. Since the {\tt MONK} code does not yet include corona--disc interaction and thermalisation of the illumination in the accretion disc, we assumed the NT73 radial temperature profile, $T_{\rm NT}(r)$, and we set the inner disc edge at ISCO. Furthermore, we assumed a BH mass $\mbh = 5\times10^7\msun$, BH spin $a=1$, accretion rate $\mdot = 0.1$, inclination $\incl= 40^\circ$ and we set temperature hardening factor to $f_{\rm col}=1$.

The colour lines in Figure~\ref{fig:MONK-KYNSED} show the {\tt MONK} results. The spectra plotted in this figure show the luminosity divided by $R_{\rm c}^2$ so that we will be able to compare them directly with the \kynsed\ results, as we discuss below. 
One can see that the resulting Comptonised emission per $R_{\rm c}^2$ gives a slightly larger normalisation for larger sizes. This is more pronounced in the case when the corona is located further away from the BH.

We used the same set-up in \kynsed\,assuming an accretion disc with $T_{\rm NT}(r)$ radial temperature profile, starting at ISCO, and without thermalisation due to the disc illumination. To achieve this, we have used the model option with external source of energy, $L_{\rm ext}$. Regarding the high energy cut-off in \kynsed, we assumed its value to be $\ecut=250\,$keV since it should be approximately 2-3 times higher than the electron temperature, see e.g., \cite{Petrucci2001}. As for the low energy cut-off, $E_0$, it is computed as an average energy of the thermal disc photons illuminating the corona, as we have described above, see e.g. eq.~(\ref{eq:E0}). Furthermore, in order to compare properly the normalization of the \kynsed\ with the MONK spectra, at each height, we chose $L_{\rm ext}$ to have the right value so that resulting size of the corona $R_{\rm c}$ is $1\rg$. In this way, the resulting \kynsed\ spectrum is identical to the one in units of power per unit area, and it can be compared directly with the MONK spectra, normalized to unit area. Note that for this comparison, the external energy feeding the corona, $L_{\rm ext}$, decreases with the height in the \kynsed\, model for the constant $R_{\rm c}=1\,\rg$, therefore the normalization of the primary X-ray spectra in Figure~\ref{fig:MONK-KYNSED} decreases with the height as opposed to the Figure~\ref{fig:height} where the energy transferred to the corona, $\Ltransf$, was kept constant.


The \kynsed\, spectra are depicted in Figure~\ref{fig:MONK-KYNSED} by the solid black line. As one can see from Figure~\ref{fig:MONK-KYNSED}, the {\tt MONK} and \kynsed\, spectra are quite similar. The \kynsed\, low energy cut-off corresponds quite well to the energy where the Comptonised emission in the {\tt MONK} spectra starts decreasing. the same is true for the high-energy cut-off in the {\tt MONK} and \kynsed\ spectra. What is perhaps even more interesting is that the normalization of the \kynsed\, spectrum is always very close to the {\tt MONK} spectra. The maximum difference between the \kynsed\ normalization and the {\tt MONK} spectra is by a factor of 1.45. This would correspond to the relative error on corona radius about 20\% ($\sqrt{1.45}$). 

To conclude, our results indicate that point source approximation used in \kynsed\, gives actually a very good approximation of the spectrum of 3D corona. The similarity between the {\tt MONK} and \kynsed\ spectra suggests that the X--ray spectrum detected by the distant observer, and the X--ray spectrum illuminating each disc element, will be the correct ones, even if the emission originates from a 3D corona, with a radius like the one that \kynsed\ computes at the end of the data fitting. Of course, the latter statement holds if the disc illumination of a point like and a 3D corona are also similar. 

Regarding the effects of the corona size with respect to the disc illumination, some extended geometries have already been studied in the past. For example, \cite{Dauser2013} studied the disc illumination by a vertically extended corona. They find that, in general, the incident radiation on the disc, as function of radius, does not differ significantly from that of a point-like source located close to the center of the jet-like structure. See for example the green-dashed and the solid red line in the left panel of their Fig. 7. There are differences, but the overall shape of the irradiation flux profile remains roughly similar. \cite{Dauser2013} only tested the vertical extension of the corona, which is different from the 3D geometry of a spherical source. However, Zhang et al. (in preparation) studied the disc irradiation in the case of a 3D spherical corona when the BH spin axis is an axis of symmetry. They found that the disc illumination pattern of the 3D corona and of a point-like source located in its center are quite similar, e.g. for the corona at height 10\rg\ the illumination profile is practically identical for the corona with its radius up to 4\rg. Thus for reasonable sizes of the corona, the illumination pattern should not differ much from the point-like approximation.


\subsection{The accretion disc density, ionisation and accretion rate.}

In the \kynsed\, model, the X-ray reflection as well as the partial absorption of the disc illumination by the corona is computed from disc re-processing  tables. As stated earlier, we used the {\tt XILLVER} tables by \cite{Garcia2013, Garcia16}. These tables have two different flavours. The first one has the high energy cut-off as one of the free input parameters while the disc density is kept constant, while in the second option the disc density is a free input parameter, while the high energy cut-off is kept at a constant value. Thus, unfortunately, we cannot change both the density and \ecut with radius in our model. 

The high energy cut-off of the incident spectrum on each disc element is shifted due to relativistic effects, the reflection and thus also the absorption changes with radius.  Further the normalisation of the Comptonised X-ray spectrum depends on the high energy cut-off, see for example Figure~\ref{fig:ecut} in the Appendix. It is for these reasons that \kynsed\ operates with the assumption of a constant disc density. 

X--ray absorption by the disc will depend on the disc ionisation (more ionised disc reflects more thus absorbs less) and the ionisation will depend on the disc density that decreases with radius further out from the center.
The ionisation also depends on the illumination that decreases with radius faster then density.
We expect most of the outer parts of the disc to be neutral and thus the absorption will not change much with the radius due to changing ionisation. Only the very inner parts of the accretion disc will be affected. 


It is possible that standard accretion discs are thermally stable until the accretion rate becomes $\sim 0.3$ of the Eddington limit or so \cite[e.g.][and references therein.]{Yuan2014}. In this case, results may not be reliable if fitting broad band SEDs of AGN with high accretion rates with any disc model. However, this may not be true for disc models, like \kynsed, when a large amount of power is transferred from the disc to the corona. \cite{Svensson1994} investigated the case when a major fraction, $f$, of the power released during the accretion process is transported to and dissipated in the corona. They showed that this has major effects on the cold disc, making it colder, more geometrically thin, denser, and having larger optical depths. One important consequence is the disappearance of the effectively optically thin zone as well as of the radiation pressure dominated zone for $f$ values close to unity (as is our case). \cite{Svensson1994} studied the case when $f$ remains constant at all radii, while $f=1$ below \rtr\, and 0 at larger radii in \kynsed. However, the radiation pressure dominated zone is usually located in the inner disc. Hence the \kynsed\ set up may result in disc solutions where gas pressure dominates at all radii, and the disc may be stable against thermal and viscous instabilities at larger accretion rates. Although these authors studied the case of spin zero only, similar results may hold for higher spins as well. This may in effect cause that, if the corona is fed by the accretion flow itself, the geometrically thin accretion disc may exist even for higher accretion rate values than the usually assumed upper limit of 0.3 or so.

\subsection{Fitting SEDs in practice}

\kynsed\ does not take into account variability in its computations. It assumes that the
spectral  properties  do  not  change  in  the  timescale  of  the  processes in action, i.e., illumination (corona to disc travel time), absorption,  thermalisation,  seed  photon  travelling  back  to  the corona,  Comptonisation  and  the  energy  transfer  from  the  disc to the coronal hot electrons. However, we know that AGN are highly variable on timescales that depend on the \mbh, e.g., for $\mbh=5\times 10^7~M_{\odot}$ the X-ray flux changes by a factor of 2 or more in just 1--2 days. The light travel time between some parts of the disc and corona may be almost 3 days if we assume their separation to be 1000~\rg. In fact, The optical (mainly) emission may be produced up to radii which may be even larger (depending on the wavelength studied). Thus, the changes in the system (notably variations in \rtr, \Ltransf, $h$, \mdot, etc.) are not seen by the corona and the disc at the same time. 
Thus at some point in time, say $t_{\rm obs}$ (i.e., the time we record the AGN flux at various bands), the disc photons received by the corona were different  than  those   emitted  by  the disc, and the Comptonised photons radiated by the corona were different than those illuminating the disc. 


The exact solution of this problem will depend on the details of the variability processes. Thus, in \kynsed\ we assume that the AGN emission process is stationary. The average (mean) X-ray spectrum radiated by the corona is sustained by the average of the energy transfer process from the accretion disc. It is produced by Comptonisation of the seed photons that correspond to the average disc energy spectrum emitted by the disc. This mean disc spectrum is the spectrum of a disc that is heated by the average mass accretion rate and the time average X--ray spectrum emitted by the corona. Therefore, the best-fit parameters that result from fitting the data with \kynsed\ should be representative of the average value of the physical parameters of the system. 

One should always be cautious when using \kynsed\ to fit broad-band SEDs. In general, \kynsed\ is not  appropriate  for  fitting data that were taken simultaneously in the X-ray, UV and optical bands, {\it if} the integration time was very short. The data sets must be taken over long time scales, so that the observed SEDs are indeed representative of the average disc spectrum that was entering into the corona, and of the average X--ray spectrum that was illuminating the disc (over the duration of the observations). The optimal duration of the observations to construct the SED depends on the amplitude of the variations in each source. For example, \kynsed\ may not be appropriate to use and fit simultaneous optical/UV/X--ray data that were taken from short observations (i.e., less than a few ksec), when the source shows significant variations on similar (or shorter) time scales. In this case, the observed X--ray spectrum may not be representative of the X--ray spectrum illuminating the whole disc, while the observed optical/UV SED may not be representative of the seed photons spectrum, when entering the corona. 

It may still be possible to use \kynsed\ to fit the data, but the SED must be constructed using only parts of the observations, and only over a limited energy range, based on some assumptions regarding the width of the disc transfer function in each band. For example, the X-ray spectrum could be extracted from the start of the observation, and the UV SED could be taken from data at the end of the observation, if the time difference between the respective data is representative of the width of the disc transfer function in the UV band. Furthermore, the UV light curve should not show significant variations on short time scales, so that the UV SED could be considered as representative of the disc photon's spectrum that entered the corona during the period that was used to extract the X--ray spectrum.

On the other hand, \kynsed\ is ideal to fit the time-averaged SEDs, which are computed using data from long, monitoring observations. As we already mentioned in the Introduction, many AGN have been observed regularly, over long periods, the last few years. One can use the data from these monitoring campaigns to construct the mean/average SED, and then fit them using \kynsed. This is what we did with the time-average SED of NGC~5548. 

\subsection{The average SED of NGC~5548}

We chose to fit the time-averaged SED of NGC~5548 which we constructed  using optical/UV/X--ray data from a long, multiwavelength observation campaign performed in 2014. We chose this object because K21b has already fitted the optical/UV time-lags that \cite{Fausnaugh16} measured using the same data set. Our results show that we can fit the broad-band, time-averaged SED assuming the same corona heights that K21a found can fit the time-lags. Our results regarding spin and accretion rate are entirely consistent with the K21b results. Therefore,  the same model, with the same model parameters, can fit both the average SED and the time-lags, which are computed from the same data sets.

Our results indicate that $\sim 45-70~\%$ of the total accretion power is transferred to the X–ray corona. This is in agreement with the results of \cite{Kubota18}, who also found that $\sim 70~\%$ of the total accretion power should be transferred to the corona. However, contrary to their results, our best-fit solution strongly favour a high spin for the BH. For such a high spin, this large power is transferred to the corona from a rather small region in the inner disc, with a radius up to $\sim 5 - 10$\rg. The radius of the corona turned out to be $\sim 12-18$\rg, which is entirely consistent with the X-ray half-light radii of quasars from recent microlensing studies of lensed systems \citep[e.g.,][]{Chartas16}.

\cite{Mathur17} fitted {\it Chandra} and {\it Swift/}XRT X-ray spectra from the same data set we analyse in this work, and required the presence of a warm Comptonizing region to account for the soft X-rays. Emission from a warm crorona was also used by \cite{Mehdipour15}, \cite{Petrucci18}, and \cite{Kubota18} to fit the broadband SED of NGC~5548 using data from a different multi-wavelength monitoring campaign, which took place in the summer of 2013.  \kynsed\ fitted the 2014,  optical/UV/X–ray spectrum very well, without assuming a warm Comptonizing region to account for the far-UV/soft X-ray emission of the source.

Arguably, the 2014 SED is missing X--ray data above $\sim 8$ keV, so the question of whether a warm corona is present in this object or not is not conclusively resolved. What is important though is that our  model,  which  is based on the assumption of X-ray illumination of the disc in the lamp-post geometry, can explain both the average SED {\it and} the optical-UV time-lags, computed using {\it simultaneous}, high quality, optical/UV/X-ray data. We also plan to study the UV/optical power spectra, assuming X--ray illumination of the disc, in more detail. We have used the response functions of K21, and we have modelled the expected PSDs. Preliminary results (Panagiotou et al., in prep.) show that the model predictions fit the NGC~5548 PSDs very well. To the best of our knowledge, ours is the only model that can explain the broadband energy spectrum, the observed UV/optical continuum time-lags, and PSDs in AGN. We note that the correlation of X-rays to UV in NGC5548 is significantly weaker than the UV-to-optical correlation, as is the case in most AGN with dense, multiwavelength monitoring data \cite[see e.g.][]{Edelson19}. At first sight, this is not in agreement with the hypothesis of disc X--ray illumination, and raises the question of whether the X-rays drive the variability at longer wavelengths. We plan to study the 
time-resolved, UV/X-ray evolution in NGC 5548 to investigate whether X--ray illumination of the disc can also account for this effect.

K21b showed that the observed UV/optical time-lags in many AGN which have been monitored regularly, over long periods, are fully consistent with the hypothesis of X–ray illuminated standard accretion discs. We plan to fit, simultaneously,  the  observed mean SED, time-lags and PSDs of these objects with  our model, to investigate whether the proposed lamp-post geometry is consistent with observations, and determine the physical parameters of these AGN as accurately as possible. Simultaneous fitting of the energy and the time-lags spectrum should be the best-way to lift the degeneracy between various physical parameters when fitting only the energy or the time-lags spectrum of the source. This is not an easy task, as we need to implement changes in the time-lags code we used in K21a and K21b (i.e., we need to add a factor of two in eq. (2) of K21a,  change the colour correction prescription, implement the option to power the corona from the power released by accretion etc.). This is beyond the scope of the current work,  Nevertheless, we plan to change the reverberation code soon, and follow this approach in the future.

\section{Acknowledgements}

\begin{acknowledgements}
MD thanks for the support from the GACR project 21-06825X and the institutional support from RVO:67985815. IEP would like to acknowledge support by the project ``Support of the international collaboration in astronomy (ASU mobility)'' with the number: CZ.$02.2.69/0.0/0.0/18-053/0016972$. ASU
mobility is co-financed by the European Union. ESK acknowledges financial support from the Centre National d’Etudes Spatiales (CNES). WZ acknowledges the support by the Strategic Pioneer Program on Space Science, Chinese Academy of Sciences through grant XDA15052100.
\end{acknowledgements}



\bibliographystyle{aa} 
\bibliography{references} 
%
%

\begin{appendix} 
\section{Plots}

\begin{figure*}
\centering
\includegraphics[width=0.6\linewidth]{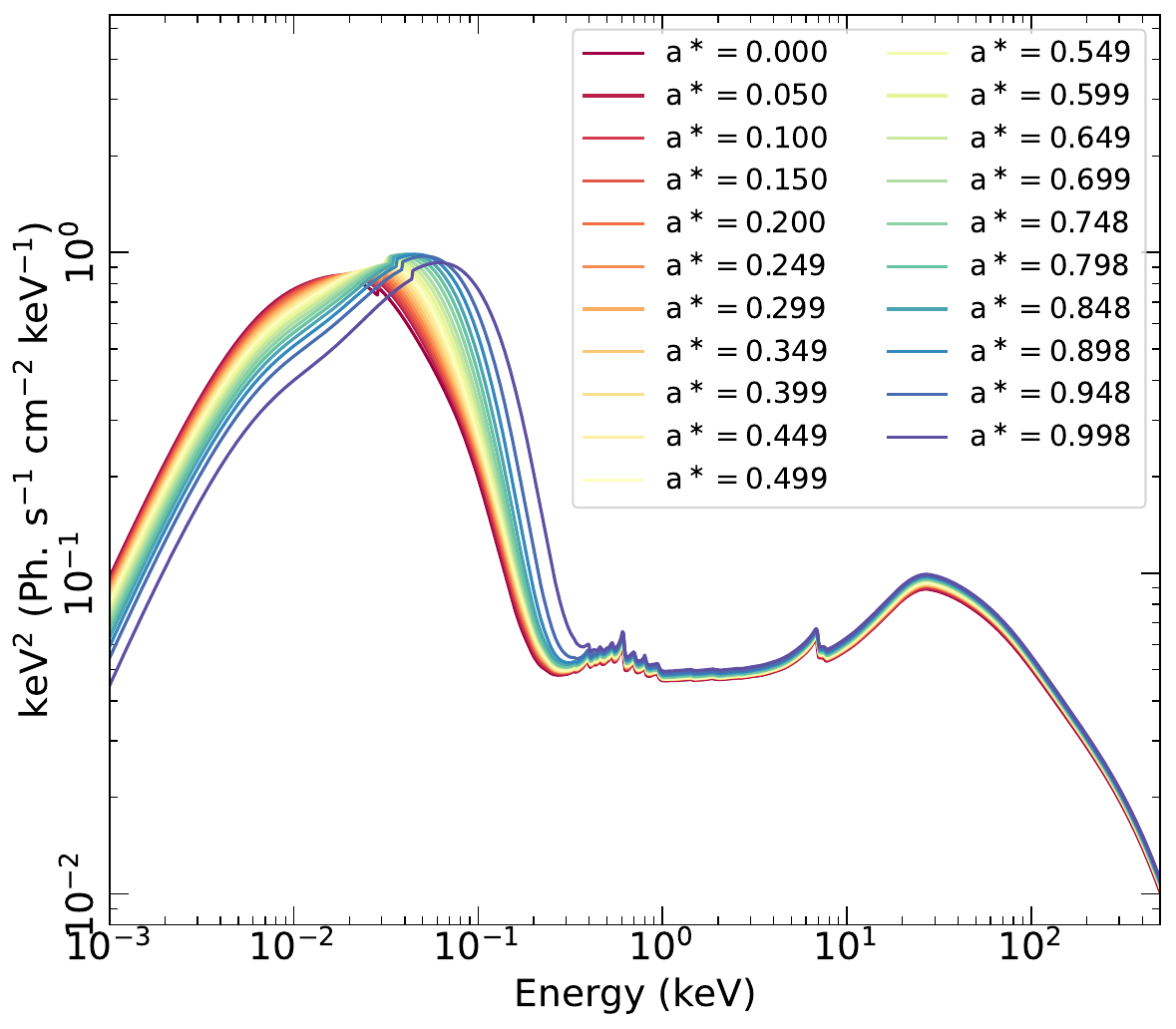}
\caption{Spectra for different BH spin values, \spin. Other parameters used: \Ltransf = 0.5, $h$ = 10 \rg, \mbh= $5\times10^7$ \msun, \mdot = 0.1, \incl = 40$^\circ$, $\Gamma = 2$ and $E_{\rm cut, obs} = 300\,$keV.}
\label{fig:spin}
\end{figure*}

\begin{figure*}
\centering
\includegraphics[width=0.95\linewidth]{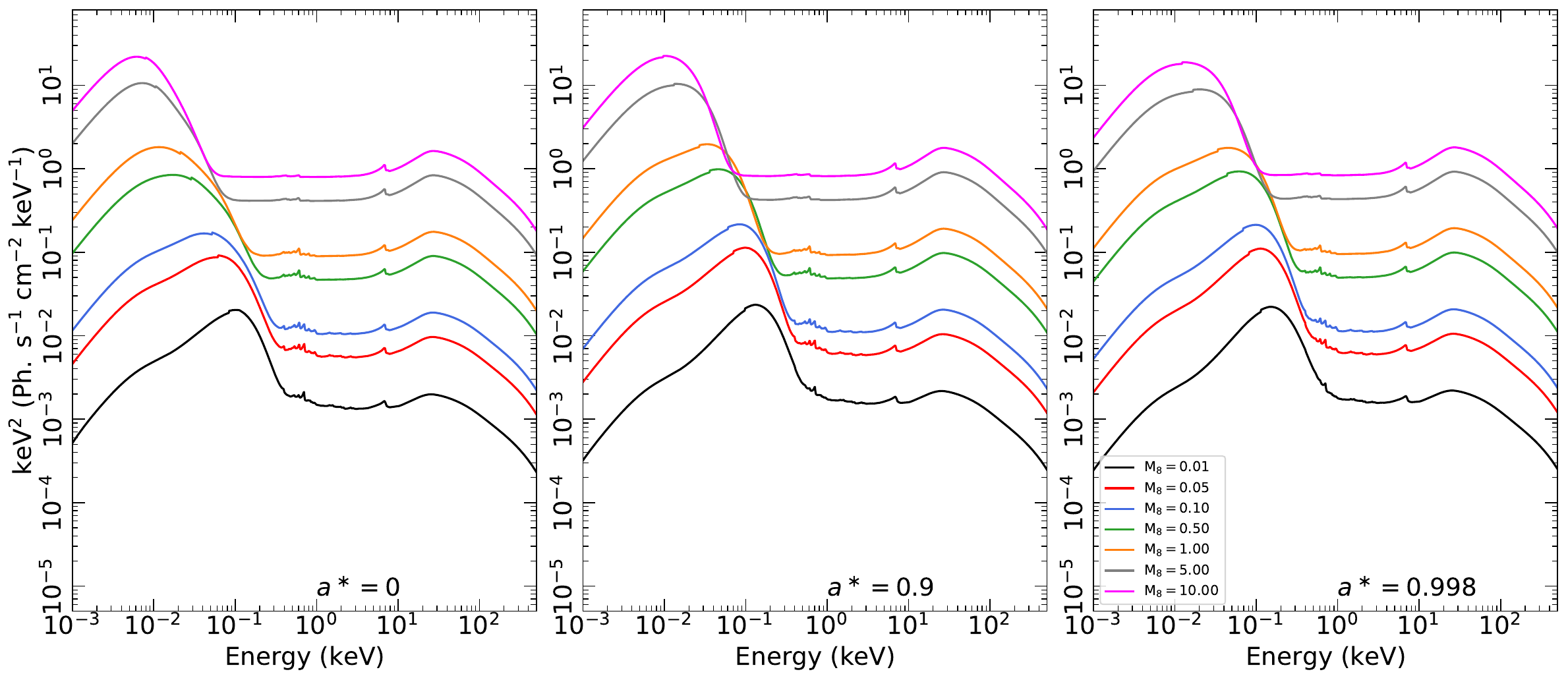}
\caption{Spectra for different BH masses, \mbh, considering spin values of \spin= 0, 0.9 and 0.998 (left, middle and right panels, respectively). Other parameters used: \Ltransf = 0.5, $h$ = 10 \rg, \mdot = 0.1, \incl = 40$^\circ$, $\Gamma = 2$ and $E_{\rm cut, obs} = 300\,$keV.}
\label{fig:mass}
\end{figure*}

\begin{figure*}
\centering
\includegraphics[width=0.95\linewidth]{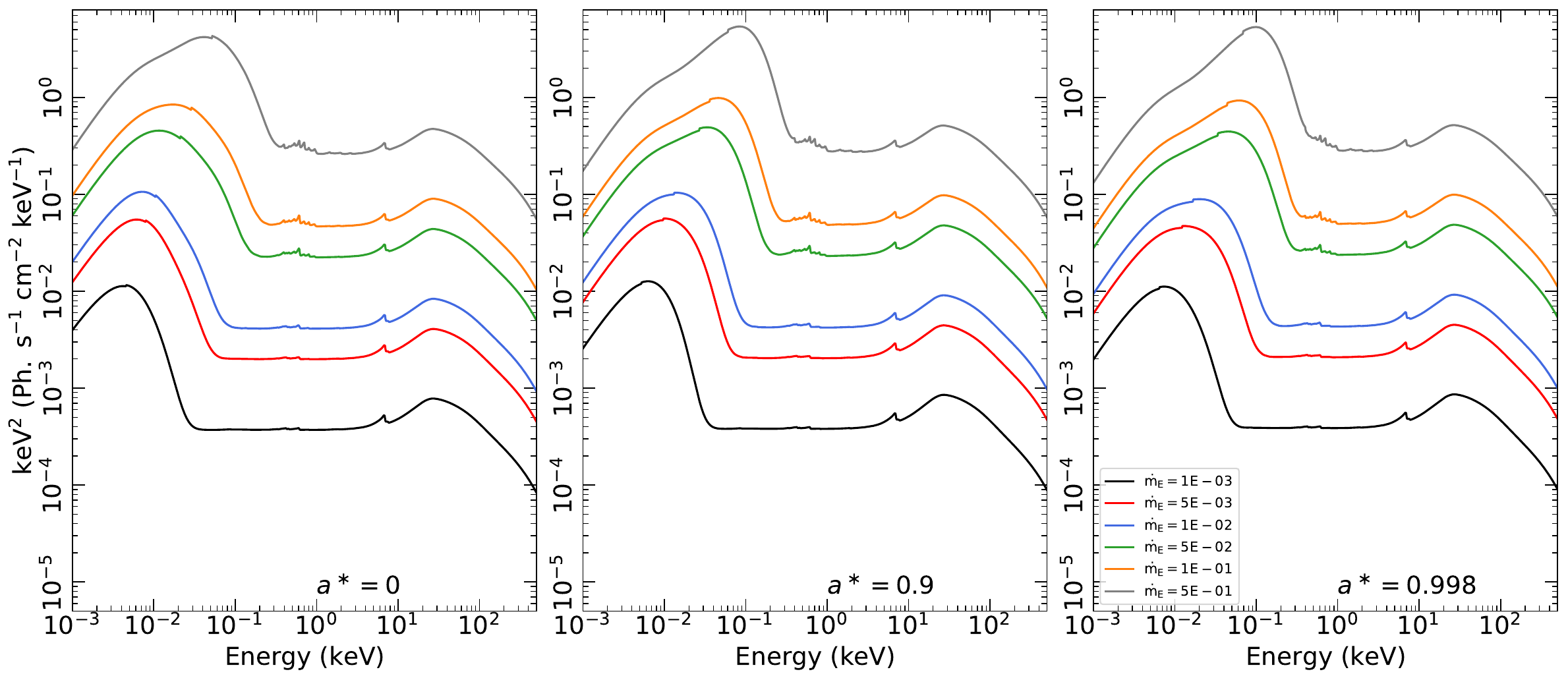}
\caption{Spectra for different accretion rates, \mdot, considering spin values of \spin= 0, 0.9 and 0.998 (left, middle and right panels, respectively). Other parameters used: \Ltransf = 0.5, $h$ = 10 \rg, \mbh= $5\times10^7$ \msun, \incl = 40$^\circ$, $\Gamma = 2$ and $E_{\rm cut, obs} = 300\,$keV.}
\label{fig:mdot}
\end{figure*}

\begin{figure*}
\centering
\includegraphics[width=0.95\linewidth]{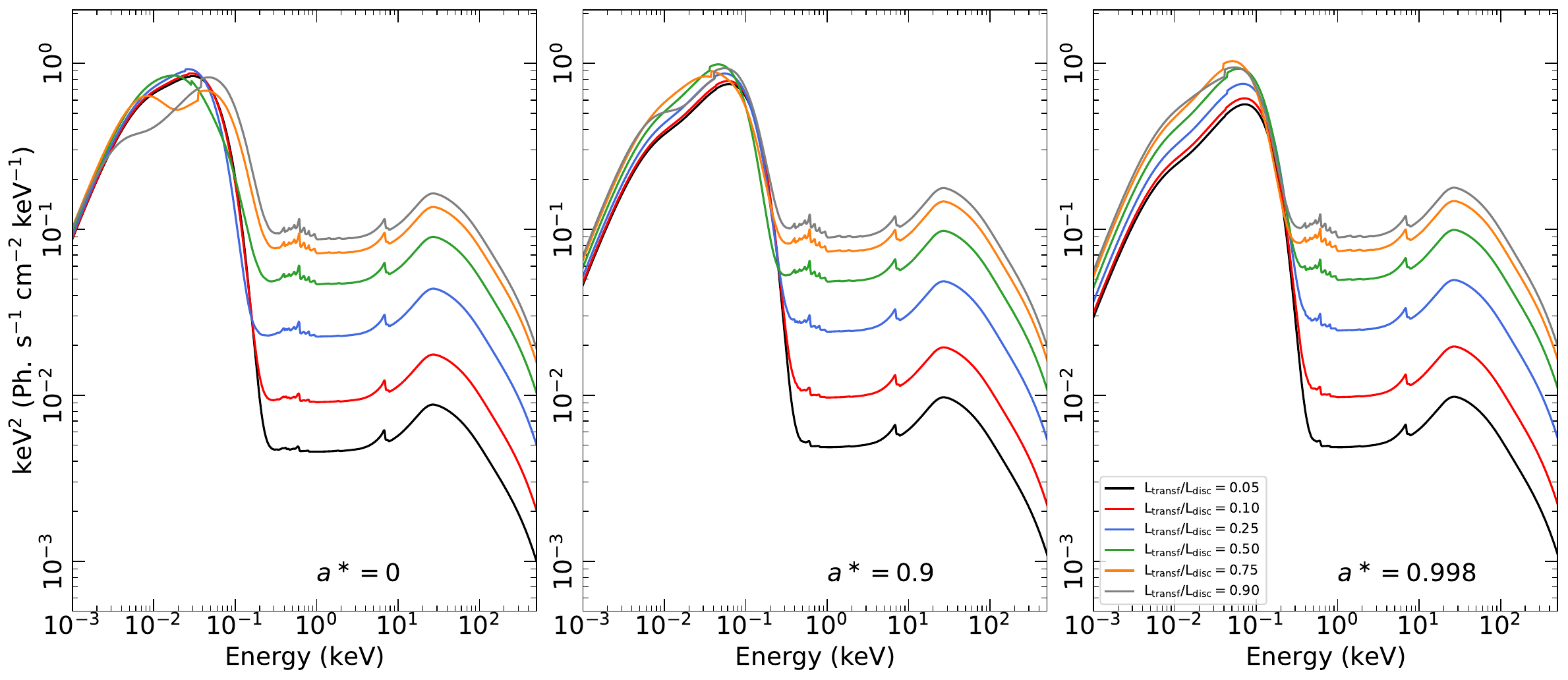}
\caption{Spectra for different values of \Ltransf, considering spin values of \spin= 0, 0.9 and 0.998 (left, middle and right panels, respectively). Other parameters used: $h$ = 10 \rg, \mbh= $5\times10^7$ \msun, \mdot = 0.1, \incl = 40$^\circ$, $\Gamma = 2$ and $E_{\rm cut, obs} = 300\,$keV.}
\label{fig:Ltransf}
\end{figure*}

\begin{figure*}
\centering
\includegraphics[width=0.95\linewidth]{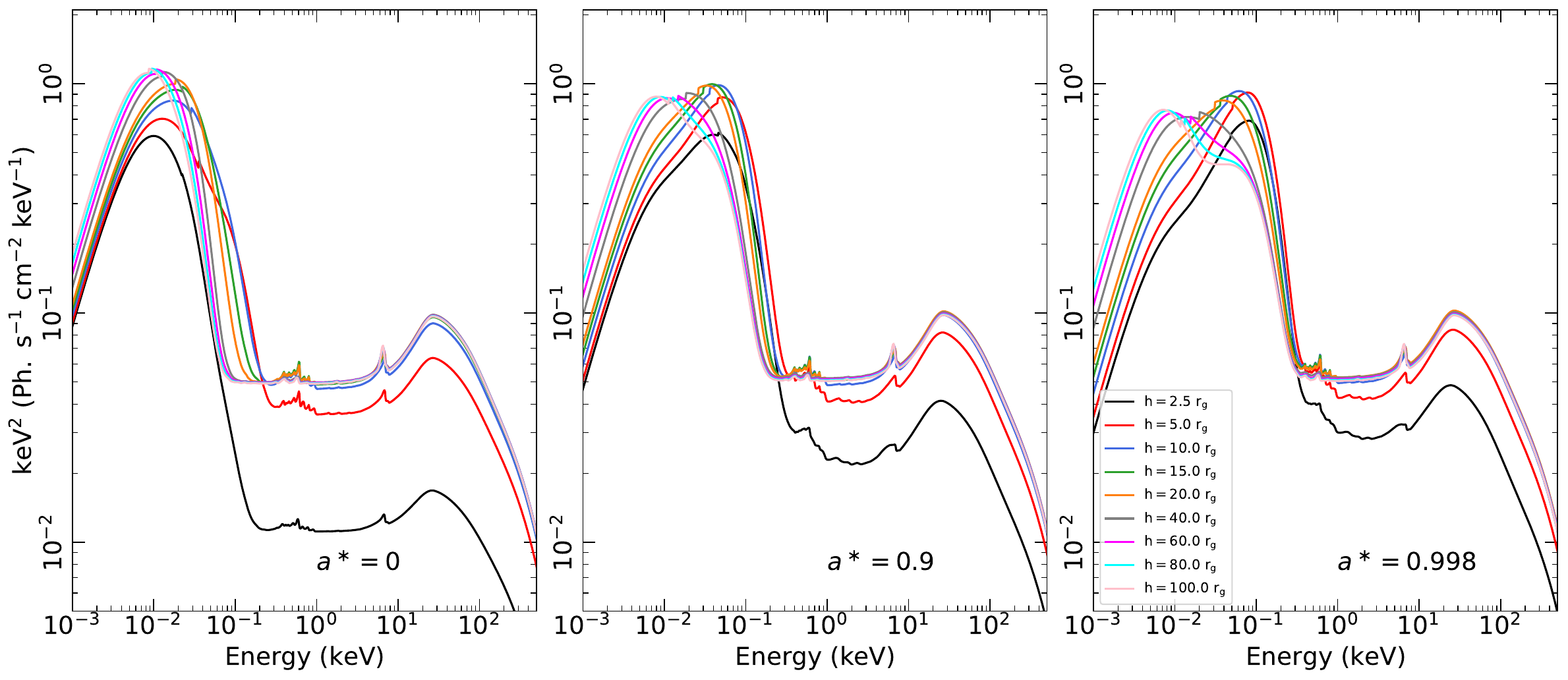}
\caption{Spectra for different corona heights, $h$, considering spin values of \spin= 0, 0.9 and 0.998 (left, middle and right panels, respectively). Other parameters used: \Ltransf = 0.5, \mbh= $5\times10^7$ \msun, \mdot = 0.1, \incl = 40$^\circ$, $\Gamma = 2$ and $E_{\rm cut, obs} = 300\,$keV.}
\label{fig:height}
\end{figure*}

\begin{figure*}
\centering
\includegraphics[width=0.95\linewidth]{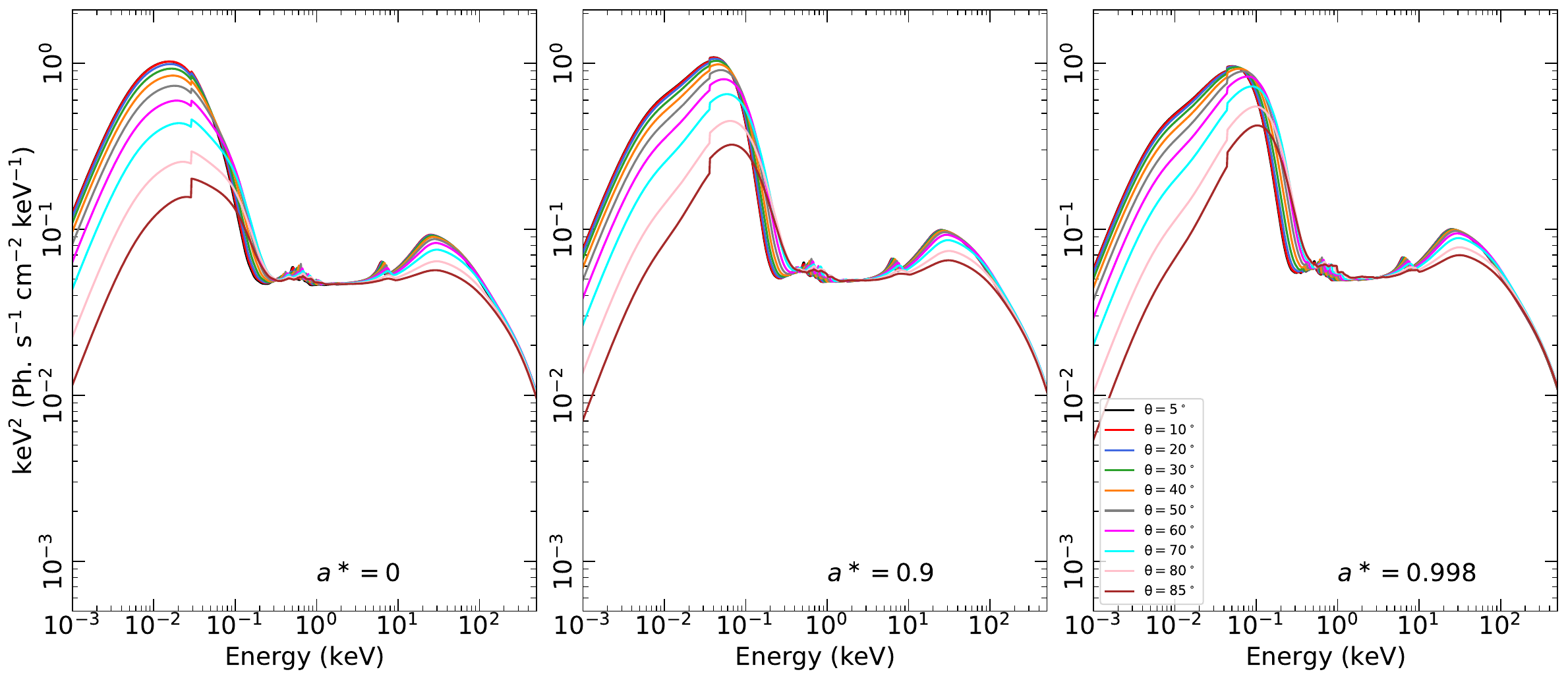}
\caption{Spectra for different observer inclinations, \incl, considering spin values of \spin= 0, 0.9 and 0.998 (left, middle and right panels, respectively). Other parameters used: \Ltransf = 0.5, $h$ = 10 \rg, \mbh= $5\times10^7$ \msun, \mdot = 0.1, $\Gamma = 2$ and $E_{\rm cut, obs} = 300\,$keV.}
\label{fig:inclination}
\end{figure*}

\begin{figure*}
\centering
\includegraphics[width=0.95\linewidth]{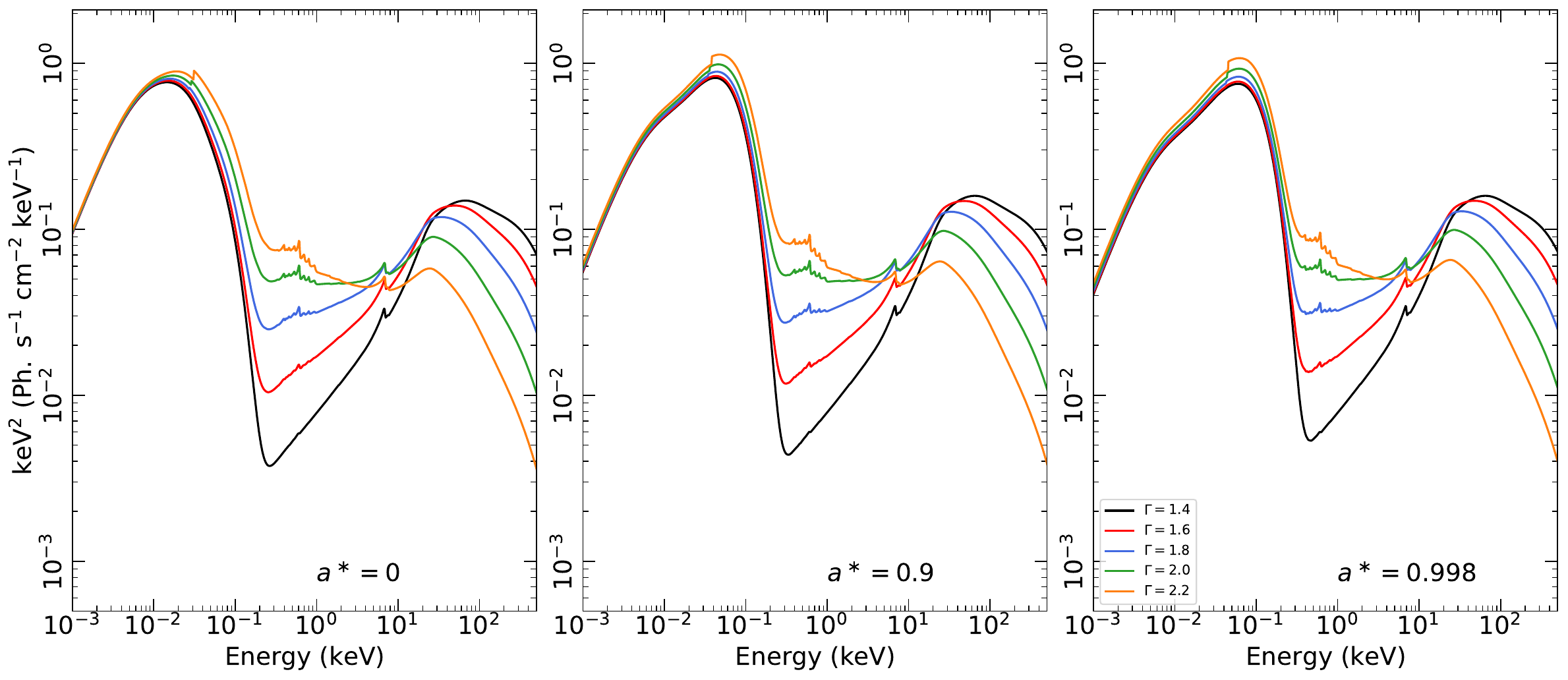}
\caption{Spectra for different photon index values of the primary power-law X-ray component, $\Gamma$, considering spin values of \spin= 0, 0.9 and 0.998 (left, middle and right panels, respectively). Other parameters used: \Ltransf = 0.5 \rg, $h$ = 10 \rg, \mbh= $5\times10^7$ \msun, \mdot = 0.1, \incl = 40$^\circ$ and $E_{\rm cut, obs} = 300\,$keV.}
\label{fig:gamma}
\end{figure*}

\begin{figure*}
\centering
\includegraphics[width=0.95\linewidth]{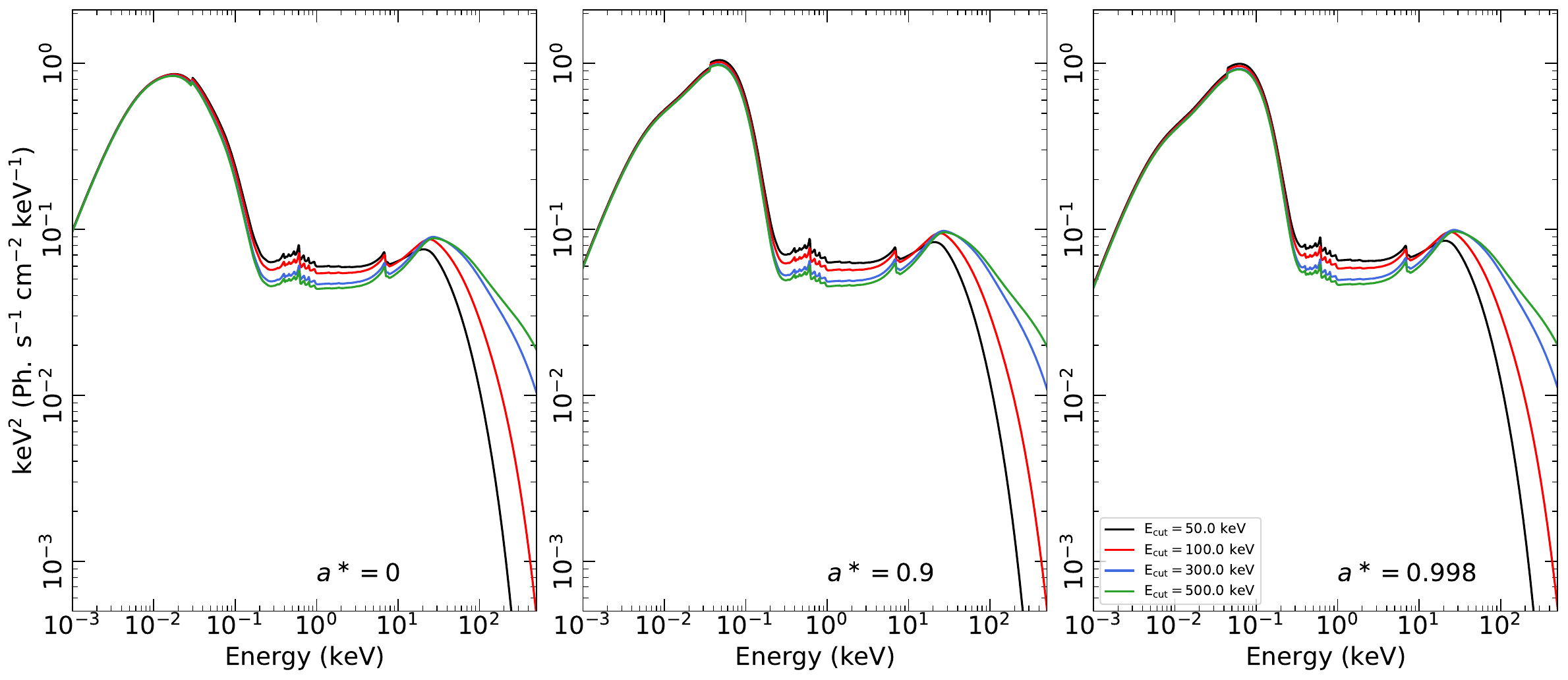}
\caption{Spectra for different high energy cut-off values of the primary power-law X-ray component, \ecut, considering spin values of \spin= 0, 0.9 and 0.998 (left, middle and right panels, respectively). Other parameters used: \Ltransf = 0.5, $h$ = 10 \rg, \mbh= $5\times10^7$ \msun, \mdot = 0.1, \incl = 40$^\circ$ and $\Gamma = 2$.}
\label{fig:ecut}
\end{figure*}

In this Section we present the plots of the SED dependence on various parameters -- spin (Figure~\ref{fig:spin}), mass (Figure~\ref{fig:mass}), accretion rate (Figure~\ref{fig:mdot}), energy transferred from the disc to the corona (Figure~\ref{fig:Ltransf}), height (Figure~\ref{fig:height}), inclination (Figures~\ref{fig:inclination}) and photon index (Figure~\ref{fig:gamma}) and high energy cut-off (Figure~\ref{fig:ecut}) of the primary power-law X-ray component.

\end{appendix}

\end{document}